\documentclass[aps,pra,reprint,superscriptaddress,letterpaper,twocolumn,longbibliography]{revtex4-1}
\usepackage[english]{babel}
\usepackage[utf8]{inputenc}
\usepackage{amsmath}
\usepackage{mathtools}
\usepackage{amssymb}
\usepackage{amsthm}
\usepackage{mathrsfs}
\usepackage{graphicx}
\usepackage{bm}
\usepackage{bbm}
\usepackage{empheq}
\usepackage{cases}
\usepackage{euscript}
\usepackage[usenames, dvipsnames, x11names]{xcolor}
\usepackage[colorlinks=true,linkcolor=SpringGreen4,citecolor=blue,urlcolor=Magenta3]{hyperref}
\usepackage{enumitem}
\newcommand{\bra}[1]{\langle #1|}
\newcommand{\ket}[1]{|#1\rangle}

\newcommand{\ketbra}[2]{\ket{#1}\!\bra{#2}}
\newcommand{\norm}[1]{\left\lVert#1\right\rVert}
\newcommand{\mm}[1]{\mathrm{#1}}
\newcommand{\abs}[1]{\left|#1\right|}

\newcommand{\di}[1]{\mathop{}\!\mathrm{d} #1}

\def \dz{\tilde{0}}
\def \tone{\tilde{1}}

\def \ua{\mathrm{a}}
\def \ub{\mathrm{b}}
\def \ug{\mathrm{g}}

\def \uq{\mathrm{q}}

\def \ue{\mathrm{e}}

\def \ud{\mathrm{d}}
\def \uq{\mathrm{q}}
\def \uI{\mathrm{I}}
\def \uG{\mathrm{G}}

\def \rd{\partial}
\def \nv{\mbox{\boldmath$n$}}
\def \sigmav{\mbox{\boldmath$\sigma$}}

\def \hrho{\hat{\rho}}

\def \hH{\hat{H}}

\def \hV{\hat{V}}
\def \hU{\hat{U}}
\def \hO{\hat{O}}

\def \hS{\hat{S}}
\def \hJ{\hat{J}}

\def \hW{\hat{W}}

\def \hP{\hat{P}}
\def \hL{\hat{L}}

\DeclareFontFamily{OT1}{pzc}{}
\DeclareFontShape{OT1}{pzc}{m}{it}{<-> s * [1.10] pzcmi7t}{}
\DeclareMathAlphabet{\mathpzc}{OT1}{pzc}{m}{it}
\def \thetat{\theta_1}
\def \thetatd{\dot{\theta}_1}

\begin{document}

\title{Accelerated adiabatic quantum gates: optimizing speed versus robustness}

\author{Hugo Ribeiro}
\affiliation{Max Planck Institute for the Science of Light, Staudtstraße 2, 91058 Erlangen, Germany}

\author{Aashish A. Clerk}
\affiliation{Institute  for  Molecular  Engineering,  University  of  Chicago, 5640  South  Ellis  Avenue,  Chicago,  Illinois  60637,  U.S.A.}

\begin{abstract}
We develop new protocols for high-fidelity single qubit gates that exploit and extend theoretical ideas for accelerated adiabatic
evolution. Our protocols are compatible with qubit architectures with highly isolated logical states, where traditional approaches
are problematic; a prime example are superconducting fluxonium qubits. By using an accelerated adiabatic protocol we can enforce
the desired adiabatic evolution while having gate times that are comparable to the inverse adiabatic energy gap (a scale that is
ultimately set by the amount of power used in the control pulses). By modelling the effects of decoherence, we explore the
tradeoff between speed and robustness that is inherent to shortcuts-to-adiabaticity approaches.
\end{abstract}

\maketitle

\section{Introduction}
\label{sec:intro}

Any approach to implementing quantum gates requires applying time-dependent control fields to a system, such that the
corresponding time-dependent Hamiltonian generates the desired unitary evolution.  Regardless of the setting, ideal gates have two
defining features: they are both robust against small imperfections in the amplitude, duration and phase of control pulses, and
they are fast. Unfortunately, typical approaches to constructing gates optimize only one of these two desired characteristics.
Schemes based on quantum adiabatic evolution (e.g.~\cite{duan2001,kis2002}) are typically extremely robust against parameter
variations~\cite{moller2008}, but suffer from extremely long protocol times.  In contrast, more conventional non-adiabatic
approaches can be extremely fast (approaching the quantum speed
limit~\cite{mandelstam1991,margolus1998,deffner2013,santos2015,pires2016,marvian2016}), but require precise tuning of control
pulses.  In a typical experimental setting, neither approach is fully optimal, as both speed and robustness are important
characteristics.

Given this, protocols that lie  between these two extremes are highly desirable.  This naturally leads one to the general approach
of shortcuts to adiabaticity (STA) \cite{demirplak2003,demirplak2005,berry2009,ibanez2012} (also known as counter-diabatic
driving).  STA are a family of techniques that allow one to mimic adiabatic evolution under some Hamiltonian $\hH_0 (t)$ using a
modified Hamiltonian $\hH_\mm{mod} (t)$, in a much shorter timescale.  STA protocols for evolving a well-defined initial state to
some prescribed well-defined final state have been discussed in many contexts, and have even recently been implemented in a
variety of experimental settings \cite{bason2011,schaff2011,zhang2013,an2016,du2016,zhou2016,koelbl2019}.  While not often
stressed, STA protocols invariably involve a tuneable trade-off between speed and robustness.  This tuneability can however be
extremely useful in a real experimental setting, where the ultimate infidelity of a gate will be influenced by both these
features.  In Ref.~\cite{zhou2016}, this tradeoff was discussed in the specific context of an accelerated STIRAP protocol
implemented in an NV center system.

\begin{figure}[t!]
	\includegraphics[width=\columnwidth]{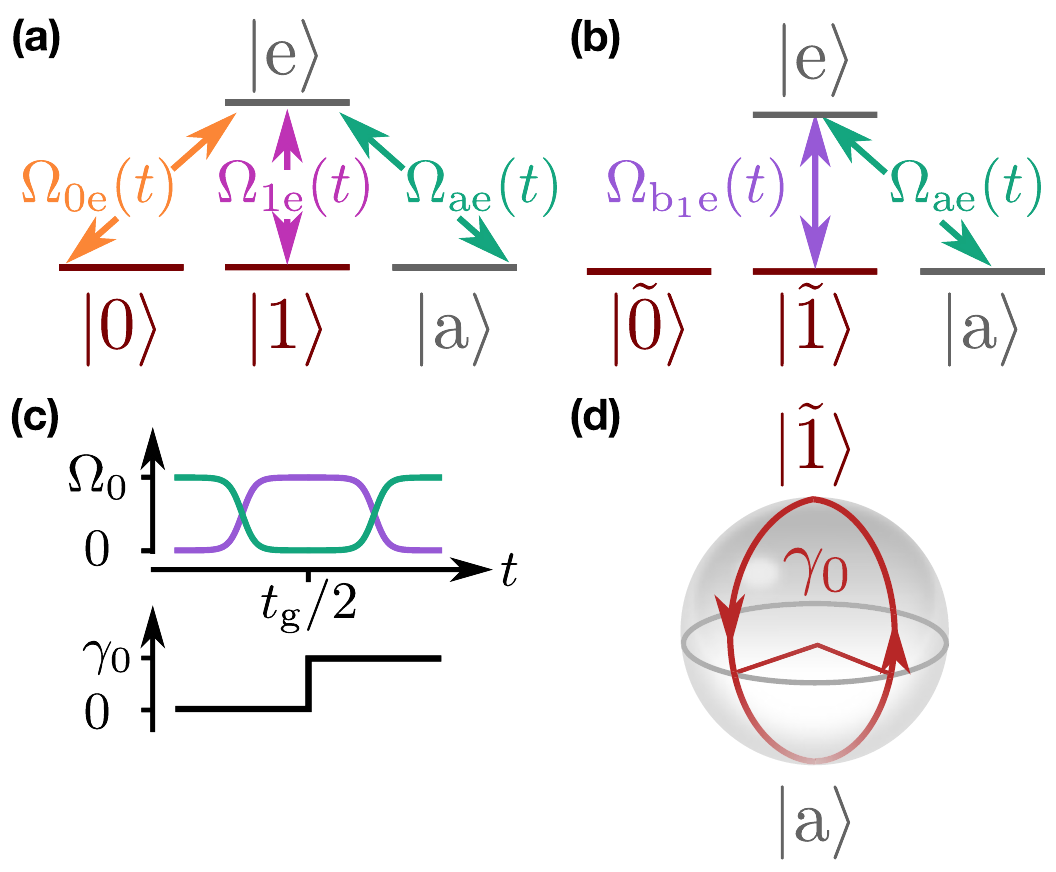}
	\caption{(Color Online) (a) Schematic representation of a tripod system. We choose $\ket{0}$ and $\ket{1}$ to encode the
		qubit states. The ground state manifold couples to an excited state $\ket{\ue}$ via control fields denoted
		$\Omega_{0\ue} (t)$, $\Omega_{1\ue} (t)$, and $\Omega_{\ua \ue} (t)$. (b) Effective $\Lambda$ system describing the
		evolution of the tripod system. The control pulses $\Omega_{\ub_1 \ue} (t)$ and $\Omega_{\ua \ue} (t)$ are chosen
		to generate a cyclic evolution based on STIRAP. (c) Example of a control sequence generating a cyclic evolution.
		The relative phase of $\Omega_{\ua \ue} (t)$ is changed instantaneously at $t=t_\ug /2$. This does not result in
		any discontinuity in the evolution since $\Omega_{\ua \ue} (t_\ug/2)=0$. (d) Geometric representation of the
		evolution of the system on the Bloch sphere. The geometric phase accumulated by $\ket{\tone}$ is equal to the
		solid angle encapsulated during the evolution.}
	\label{fig:fig01} 
\end{figure}
					
In this paper, we investigate the use of STA techniques to accelerate well-known adiabatic quantum gates based on a tripod level
configuration, where three ``ground state'' levels all interact controllably with a single ``excited state'' level. Such schemes
can find direct application in a variety of systems, including trapped ion qubits~\cite{duan2001,srinivas2019} as well as
superconducting qubits~\cite{manucharyan2009,earnest2018}. Accelerating a quantum gate is a more challenging problem than simply
accelerating an adiabatic evolution with a single, well-defined initial state, as now one is interested in a manifold of possible
initial states.  In the case of a unique initial state, the (global) phase accumulated during the evolution is of no importance
and it can consequently differ between the adiabatic and accelerated protocol. In stark contrast, when generating a quantum gate,
the accumulated phases are of utmost importance.  This is problematic, as standard STA techniques are not designed to preserve
dynamical or geometric phases generated by adiabatic evolution.  Despite this difference, we show that the superadiabatic
transitionless driving (SATD) scheme developed in Ref.~\cite{baksic2016} to accelerate STIRAP-style quantum state transfer can be
used to accelerate tripod-based adiabatic quantum gates.  We also study in detail the tradeoffs entailed with using an accelerated
protocol:~while the protocol time can be dramatically reduced, one necessarily also becomes less tolerant of parameter variations,
and also more sensitive to dissipative effects originating with the lossy excited level.  The understanding we develop will allow
one to design an optimally-constructed accelerated protocol for a given set of experimental parameters. 

In contrast to other schemes~\cite{sjoqvist2012,sjoqvist2016,abdumalikov2013,zhou2017} that generate non-adiabatic geometric
gates, ours is purely geometric and does not rely on accumulating specific dynamical phases.  This also distinguishes our work
from the recent experiment by Yan \textit{et al.}~\cite{yan2019}, where an accelerated geometric gate is only obtained if a
dynamical phase is cancelled by applying a $\pi$-pulse. We also note that while Ref.~\cite{liu2017} straightforwardly applied the
\emph{dressed state} technique of Ref.~\cite{baksic2016} to accelerate an adiabatic gate, they did not consider the potential
difficulties associated with this procedure (stemming from STA-induced modification of phases).  

\section{Geometric Gates in a Tripod System} 

\subsection{Basic double-STIRAP protocol}
\label{subsec:AdiabaticGate}

We start by reviewing how geometric qubit gates can be implemented in a four-level tripod system [see Fig.~\ref{fig:fig01}~(a)].
Our discussion complements existing literature~\cite{duan2001,kis2002} by providing a thorough discussion on how non-adiabatic
errors deteriorate the performance of such gates.  The system consists of three ground state levels ($\ket{0}$, $\ket{1}$, and
$\ket{\ua}$), each of which is controllably and resonantly coupled to a common excited state $\ket{\ue}$.  The system Hamiltonian
is 
\begin{equation}
	\hH (t) = \frac{1}{2} \left[ \Omega_{0\ue} (t) \ketbra{0}{\ue} + \Omega_{1\ue} (t) \ketbra{1}{\ue} + \Omega_{\ua \ue} (t)
	\ketbra{\ua}{\ue} + \mm{H.c.}\right],
	\label{eq:HTripod}
\end{equation}
where $\Omega_{i \ue} (t)$ $(i=0,1,\ua)$ denotes the complex envelope of each control field.  

We will use the ground states $\ket{0}$ and $\ket{1}$ to encode a logical qubit state.  It allows one to use highly isolated
states as qubit levels, thus potentially enabling long coherence times.  This kind of situation can be realized in a variety of
experimental platforms, e.g.~in superconducting fluxonium qubits~\cite{manucharyan2009,earnest2018}.  

We next parametrize the control fields, assuming only that they are chosen to keep the instantaneous eigenvalues of $\hH (t)$
independent of time:
\begin{equation}
	\begin{aligned}
		\Omega_{0 \ue} (t) &= \Omega_0 \cos(\alpha) \sin[\theta(t)],\\
		\Omega_{1 \ue} (t) &= \Omega_0 \sin(\alpha) \sin[\theta(t)] e^{i \beta}, \\
		\Omega_{\ua \ue} (t) &= \Omega_0 \cos[\theta(t)] e^{i \gamma (t)}.
	\end{aligned}
	\label{eq:EnvCtrl}
\end{equation}
$\Omega_0$ determines the overall scale for the control fields [and the size of the energy gap of $\hH (t)$], while the angles
$\theta$ and $\alpha$ determine their relative magnitudes.  The angles $\beta$ and $\gamma$ control the relative phases between
control fields.  For reasons that will become clear, we consider in what follows protocols where $\alpha$ and $\beta$ are
time-independent. 

Diagonalizing the instantaneous Hamiltonian $\hH (t)$, one finds that it always possesses two zero-energy eigenstates that are
orthogonal to $\ket{\ue}$. States in this ``dark state'' manifold are ideally suited for geometric gates, as they will never
acquire dynamical phases. Further, there is always a unique dark state that is a superposition of qubit states only, namely
\begin{equation}
	\ket{\dz} = \sin(\alpha)\ket{0} - \exp(i \beta)\cos(\alpha)\ket{1}.
	\label{eq:0Tilde}
\end{equation}
This state does not depend on time. The orthogonal qubit-only state is
\begin{equation}
	\ket{\tone} = \cos(\alpha) \ket{0} + \exp(i \beta)\sin(\alpha)\ket{1},
	\label{eq:1Tilde}
\end{equation}
and in general is not an instantaneous eigenstate of $\hH (t)$.

Writing $\hH (t)$ in terms of these new qubit basis states yields
\begin{equation} 
	\hH(t) =\frac{1}{2} \left[\Omega_{\tone \ue} (t)\ketbra{\tone}{\ue} + \Omega_{\ua \ue} (t) \ketbra{\ua}{\ue} 
	+ \mm{H.c.}\right],
\label{eq:HLambda} 
\end{equation}
where $\Omega_{\tone \ue} (t) = \Omega_0 \sin[\theta(t)]$.  We see that the qubit state $\ket{\dz}$ is completely decoupled,
whereas the qubit state $\ket{\tone}$ forms a three-level $\Lambda$-system~\cite{bergmann1998,vitanov2017} with the states
$\ket{\ua}$ and $\ket{\ue}$ [see Fig.~\ref{fig:fig01}~(b)].  One can now use well-known STIRAP
protocols~\cite{bergmann1998,vitanov2017} to adiabatically manipulate these states. In particular, using an appropriate
double-STIRAP protocol we can engineer a cyclic evolution, such that the qubit state $\ket{\tone}$ acquires a purely geometric Berry
phase~\cite{pancharatnam1956,berry1984}.  This will form the basis of our adiabatic single qubit gate (as first suggested in Refs.
\cite{duan2001,kis2002}).

To understand the double-STIRAP protocol, we first list the remaining instantaneous eigenstates of $\hH(t)$.  In addition to the
zero energy qubit dark state $\ket{\dz}$ [c.f.~Eq.~\eqref{eq:0Tilde}], $\hat{H}(t)$ in Eq.~\eqref{eq:HLambda} also has a
second, orthogonal zero energy dark state
\begin{equation}
	\ket{\ud_2 (t)} =\cos[\theta(t)]\ket{\tone} - e^{i \gamma(t)} \sin[\theta(t)]\ket{\ua}.
    \label{eq:dark2}
\end{equation}
as well as two non-zero energy eigenstates:
\begin{equation}
	\ket{\ub_\pm (t)} = \frac{1}{\sqrt{2}} \left( \pm \sin[\theta(t)]\ket{\tone} \pm e^{i \gamma (t)} \cos[\theta(t)] \ket{\ua} + \ket{\ue} \right)
	\label{eq:brightSts}
\end{equation}
with instantaneous energies $\pm \Omega_0/2$.

The double-STIRAP protocol involves adiabatically evolving the dark state $\ket{\ud_2(t)}$ from being purely $\ket{\tone}$ at $t=0$,
to being $\ket{\ua}$ at $t=t_\ug/2$, and then back to being $\ket{\tone}$ at the final time $t = t_\ug$.  This can be accomplished
by choosing  
\begin{equation}
	\theta (t) =
	\begin{dcases}
		\frac{\pi}{2} P(t) &0 \leq t < \frac{t_\ug}{2} \\
		\frac{\pi}{2} \left[1 - P \left(t-\frac{t_\ug}{2}\right) \right] &\frac{t_\ug}{2} \leq t \leq t_\ug,
	\end{dcases}
	\label{eq:theta}
\end{equation}
where $P(t)$ is a monotonic function varying between $P(0) = 0$ and $P(t_\ug /2) =1$.  The following form for $P(t)$ is
particularly effective:
\begin{equation}
	P(t) = 6 \left( \frac{2 t}{t_\ug} \right)^5 - 15 \left( \frac{2 t}{t_\ug} \right)^4 + 10 \left(
	\frac{2 t}{t_\ug} \right)^3.
	\label{eq:polyangle}
\end{equation}
This choice gives a smooth turn on and turn off of the pulses, i.e.~it satisfies $\dot{\theta} (0) = \dot{\theta} (t_\ug /2) =
\dot{\theta} (t_\ug) = \ddot{\theta} (0) = \ddot{\theta} (t_\ug /2) = \ddot{\theta} (t_\ug) = 0$. Note that at this stage, we do
not specify the time-dependence of the relative phase $\gamma(t)$; as we will see, $\gamma(t)$ will determine the geometric phase
acquired by $\ket{\tone}$. We assume for clarity that the control field  $\Omega_{\ua \ue} (t)$ is non-zero at $t=0$.  This is not
restrictive.  Even if one includes a finite turn-on time for this field, the additional resulting dynamics {\it only} effects the
states $\{\ket{\ua},\ket{\ue}\}$  (i.e.~ the auxiliary subspace.  As shown in what follows, this does not hinder the realization
of our geometric gate, as this gate is not contingent on any special preparation of the  auxiliary subspace.

The system dynamics are best analyzed in the adiabatic frame. The frame-change operator is the time-dependent unitary
$\hS_\mm{ad}(t)$ that diagonalizes $\hH (t)$ at each instant:
\begin{equation}
	\hS_\mm{ad} (t) = \ketbra{\dz}{\dz} + \ketbra{\ud_2 (t)}{\ud_2} + \ketbra{\ub_- (t)}{\ub_-} + \ketbra{\ub_+
	(t)}{\ub_+}.
	\label{eq:SadLambda}
\end{equation}
In the adiabatic frame, we have
\begin{equation}
	\begin{aligned}
		\hH_{\mm{ad}} (t) &= \hS_\mm{ad}^\dag (t) \hH (t) \hS_\mm{ad} (t) - i \hS_\mm{ad}^\dag (t) \rd_t
		\hS_\mm{ad} (t) \\
		&= \hH_0 (t) + \hV_\mm{err} (t),
	\end{aligned}
	\label{eq:HLambdaAdFrame}
\end{equation}
where
\begin{equation}
	\begin{aligned}
		\hH_0 (t) &= -\frac{\Omega_0}{2} \left(\ketbra{\ub_-}{\ub_-} - \ketbra{\ub_+}{\ub_+}\right) + \dot{\gamma} (t)
		\sin[\theta(t)]^2 \ketbra{\ud_2}{\ud_2}\\ &\phantom{={}} + \frac{1}{2} \dot{\gamma} (t) \cos[\theta(t)]^2 \left(
		\ketbra{\ub_-}{\ub_-}+\ketbra{\ub_+}{\ub_+}
		\right)
	\end{aligned}
	\label{eq:H0Lambda}
\end{equation}
is a diagonal operator which generates the desired adiabatic evolution.  In contrast, $\hV_\mm{err} (t)$ describes non-adiabatic
errors in the evolution: 
\begin{equation}
	\begin{aligned}
		\hV_\mm{err} (t) &= \frac{\dot{\theta}(t)}{\sqrt{2}} ( i \ketbra{\ud_2}{\ub_-} - i
		\ketbra{\ud_2}{\ub_+} + \mm{H.c.}) \\ 
		&\phantom{={}}  
		+ \frac{\dot{\gamma} (t)}{2} \left[
		-\cos[\theta(t)]^2 \ketbra{\ub_+}{\ub_-}
		+ \frac{\sin[2\theta (t)]}{\sqrt{2}} \ketbra{\ud_2}{\ub_-} \right. \\
		&\phantom{={}}
		\left. - \frac{\sin[2\theta (t)]}{\sqrt{2}} \ketbra{\ud_2}{\ub_+}  + \mm{H.c.}\right].
	\end{aligned}
	\label{eq:VerrLambda}
\end{equation}
Equation~\eqref{eq:VerrLambda} differs from the non-adiabatic Hamiltonian derived in Ref.~\cite{baksic2016} because the latter
work did not consider STIRAP with time-dependent relative phases.

If one now assumes that we are in the adiabatic limit,  i.e. $2\dot{\theta} (t) / \Omega_0 \to 0$ and $2\dot{\gamma} (t) /
\Omega_0 \to 0$, then we can ignore $\hV_\mm{err} (t)$, and the unitary operator describing the evolution is 
\begin{equation}
	\begin{aligned}
		\hU_\mm{ad} (t) &= \ketbra{\dz}{\dz} + e^{-i \gamma_0} \ketbra{\ud_2}{\ud_2} \\
		&\phantom{={}} 
		+ e^{i \left(\frac{\Omega_0}{2}t - \gamma_1 \right)}\ketbra{\ub_-}{\ub_-} 
		+ e^{-i \left(\frac{\Omega_0}{2}t + \gamma_1\right)}\ketbra{\ub_+}{\ub_+} \\
		&\phantom{={}} 
		+ \mathcal{O}\left[\frac{\dot{\theta} (t)}{\Omega_0}, \frac{\dot{\gamma} (t)}{\Omega_0}\right],
	\end{aligned}
	\label{eq:Uad0Lambda}
\end{equation}
where
\begin{equation}
	\begin{aligned}
		\gamma_0 &= \int_0^{t_\ug} \di{t_1} \sin[\theta(t_1)]^2 \dot{\gamma} (t_1),\\
		\gamma_1 &= \frac{1}{2}\int_0^{t_\ug} \di{t_1} \cos[\theta(t_1)]^2 \dot{\gamma} (t_1)
	\end{aligned}
	\label{eq:GeomPhaseGen}
\end{equation}
are the geometric phases accumulated by the dark and bright states, respectively.

Before proceeding, we note that there is an extremely simple choice for the relative control field phase $\gamma(t)$ that, despite
first appearances, is compatible with adiabatic evolution.  Namely, one can use
\begin{equation}
	\gamma (t) = \gamma_0 \Theta\left(t-\frac{t_\ug}{2}\right),
	\label{eq:PhaseAE}
\end{equation}
where $\Theta (t)$ denotes the Heaviside step function.  Despite the discontinuity at $t = t_\ug/2$, there is no issue with
adiabaticity. Note that our chosen pulse shapes [c.f.~Eq.~\eqref{eq:EnvCtrl}] satisfy $\theta(t_\ug/2) = \pi/2$, implying that the
phase $\gamma(t)$ is not well defined at this time; this allows the jump in Eq.~\eqref{eq:PhaseAE}. For a more geometric picture,
note that this type of control results in a trajectory on the Bloch sphere that resembles a citrus wedge, as depicted in
Fig.~\ref{fig:fig01}~(d). Finally we emphazise that this choice leads to $\gamma_1 = 0$ [c.f. Eq.~\eqref{eq:GeomPhaseGen}].

We can express Eq.~\eqref{eq:Uad0Lambda} in the (time-independent) lab frame via the transformation $\hU_\mm{ad} (t) = \hS_\mm{ad}
(t) \hU_{\Lambda,\mm{ad}} (t) \hS_\mm{ad}^\dag (0)$. Using for $\gamma (t)$ Eq.~\eqref{eq:PhaseAE}, we find at the final time
$t=t_\ug$
\begin{equation}
	\hU_{\uG,\mm{ad}} = e^{-i \frac{\gamma_0}{2}} e^{-i \frac{\gamma_0}{2} \nv \cdot \hat{\sigmav}_{01}} \oplus e^{i
	\frac{\gamma_0}{2}} e^{i \frac{\varphi_\mm{ad}}{2} \nv_{\mm{ad}} \cdot \hat{\sigmav}_{\ua \ue}}.
	\label{eq:GeomGate}
\end{equation}
Here, $\hat{\sigmav}_{01} = (\ketbra{0}{1}+\mm{H.c.},\,-i\ketbra{0}{1}+ \mm{H.c.},\,\ketbra{0}{0} - \ketbra{1}{1})$  and
$\hat{\sigmav}_{\ua \ue} = (\ketbra{\ua}{\ue}+\mm{H.c.},\,-i\ketbra{\ua}{\ue}+ \mm{H.c.},\,\ketbra{\ua}{\ua} - \ketbra{\ue}{\ue})$
denote a vector of Pauli matrices. We have further defined the unit vectors
\begin{equation}
\begin{aligned}
    \nv & =
        [\sin(2\alpha) \cos(\beta), 
	\sin(2\alpha) \sin(\beta), \cos(2\alpha)], \\
    \nv_\mm{ad} &= 
    	\frac{\sin \left(\frac{\Omega_0 t_\ug}{2}\right)}
	{\sin \left( \frac{\varphi_\mm{ad}}{2} \right)} \sin \left(\frac{\gamma_0}{2} \right)
        \left[-\cot \left(\frac{\gamma_0}{2}\right),
        1, 
        \cot \left(\frac{\Omega_0 t_\ug}{2} \right)
	\right],
\end{aligned}
\label{eq:UnitVecRot}
\end{equation}
and the rotation angle 
\begin{equation}
	\varphi_\mm{ad} = 2 \arccos\left[\cos\left(\frac{\gamma_0}{2}\right) \cos\left(\frac{\Omega_0 t_\ug}{2}\right)\right].
	\label{eq:AdRotAng}
\end{equation}

Equation~\eqref{eq:GeomGate} shows clearly that at $t=t_\ug$ the qubit subspace is decoupled from that of the two auxiliary
levels. The evolution in the qubit subspace is a simple rotation. The rotation axis $\nv$ is controlled by the static pulse
parameters $\alpha, \beta$, whereas the rotation angle $\gamma_0$ is a geometric phase. We thus have a purely geometric arbitrary
single qubit gate. We stress that having a gate that acts independently on the qubit and auxiliary subspaces is crucial:~it allows
a qubit gate to be performed {\it without} first having to prepare the state of the auxiliary levels.    

\subsection{Non-adiabatic errors}

Our goal is to accelerate the adiabatic gate described above.  As a first step, we need to understand the effects of non-adiabatic
errors that occur when the protocol time is not infinitely slow compared to the inverse instantaneous energy gap $2/\Omega_0$ of
$\hH (t)$.

We can calculate non-adiabatic corrections to the evolution perturbatively in $\hV_\mm{err} (t)$ [c.f.~Eq.~\eqref{eq:VerrLambda}]
using a Magnus expansion~\cite{magnus1954}. Using $\theta (t)$ as defined in Eq.~\eqref{eq:theta} with $P (t)$ given by
Eq.~\eqref{eq:polyangle} and $\gamma (t)$ as defined in Eq.~\eqref{eq:PhaseAE}, we find to leading order (see
Appendix~\ref{sec:appendixA})
\begin{equation}
	\hU_\uG = 
	\left(
	e^{-i \frac{\gamma_0}{2}} e^{-i \frac{\gamma_0}{2} \nv \cdot \hat{\sigmav}} 
	\oplus
	e^{i
	\frac{\gamma_0}{2}} e^{i \frac{\varphi_\mm{na}}{2} \nv_{\mm{na}} \cdot \hat{\sigmav}_{\ua \ue}}
	\right) 
	+
	\mathcal{O}\left[\frac{1}{(\Omega_0 t_\ug)^3}\right],
	\label{eq:GeomGateNonAdCorr}
\end{equation}
with 
\begin{equation}
	\begin{aligned}
		\nv_\mm{na} &= \frac{\sin \left[\varphi_0 (t_\ug)\right]}
		{\sin \left( \frac{\varphi_\mm{na}}{2} \right)} \sin \left(\frac{\gamma_0}{2} \right)
        	\left\{-\cot \left(\frac{\gamma_0}{2}\right),
		1, \cot \left[\varphi_0 (t_\ug)\right] \right\}, \\
		\varphi_\mm{na} &= 2 \arccos\left\{\cos\left(\frac{\gamma_0}{2}\right) \cos\left[\varphi_0
		(t_\ug)\right]\right\},\\
		\varphi_0 (t_\ug) &= \frac{\Omega_0 t_\ug}{2} + \frac{10 \pi^2}{7\Omega_0t_\ug}.
\end{aligned}
\label{eq:NaRot}
\end{equation}
Comparing against Eqs.~\eqref{eq:GeomGate} and \eqref{eq:GeomGateNonAdCorr}, we see that to leading order, non-adiabatic errors do
not change the nature of the gate: we still have a pure geometric operation on the qubit subspace and the latter is still
decoupled from the auxiliary subspace. Non-adiabatic errors only change the rotation performed on the auxiliary levels.  The
favourable scaling here is a direct consequence of our smooth choice for $\theta (t)$ [c.f.~Eq.~\eqref{eq:theta}], whose
derivatives vanish at the protocol start and end; this corresponds to the ``boundary cancellation method'' for non-adiabatic error
suppression discussed in Refs.~\cite{lidar2009,rezakhani2010,wiebe2012}.

A more useful measure for quantifying the impact of non-adiabatic errors is given by the state-averaged fidelity of the gate~\cite{pedersen2007}
\begin{equation}
	\bar{F} = \frac{\mm{Tr}\left[\hO \hO^\dag \right] + \abs{\mm{Tr}\left[ \hO\right]}^2}{d(d+1)},
	\label{eq:avgF}
\end{equation}
where $\mm{Tr}[\cdot]$ denotes the trace operation, $d=4$ is the dimension of the Hilbert space, and $\hO = \hU_{\uG,\mm{ad}}^\dag
\hU_\uG$ where $\hU_{\uG,\mm{ad}}$ is given in Eq.~\eqref{eq:GeomGate} and $\hU_\uG$ is the unitary operator generated by
Eq.~\eqref{eq:HTripod} evaluated at $t=t_\ug$. Using the approximate $\hU_\uG$ given in Eq.~\eqref{eq:GeomGateNonAdCorr}, we find 
\begin{equation}
	\bar{F} = 1 - \frac{40 \pi^4}{49 \left(\Omega_0 t_\ug\right)^2} + 
	\mathcal{O}\left[\frac{1}{(\Omega_0 t_\ug)^4}\right].
\label{eq:AvgFidFull}
\end{equation}

\begin{figure}[t!]
	\includegraphics[width=\columnwidth]{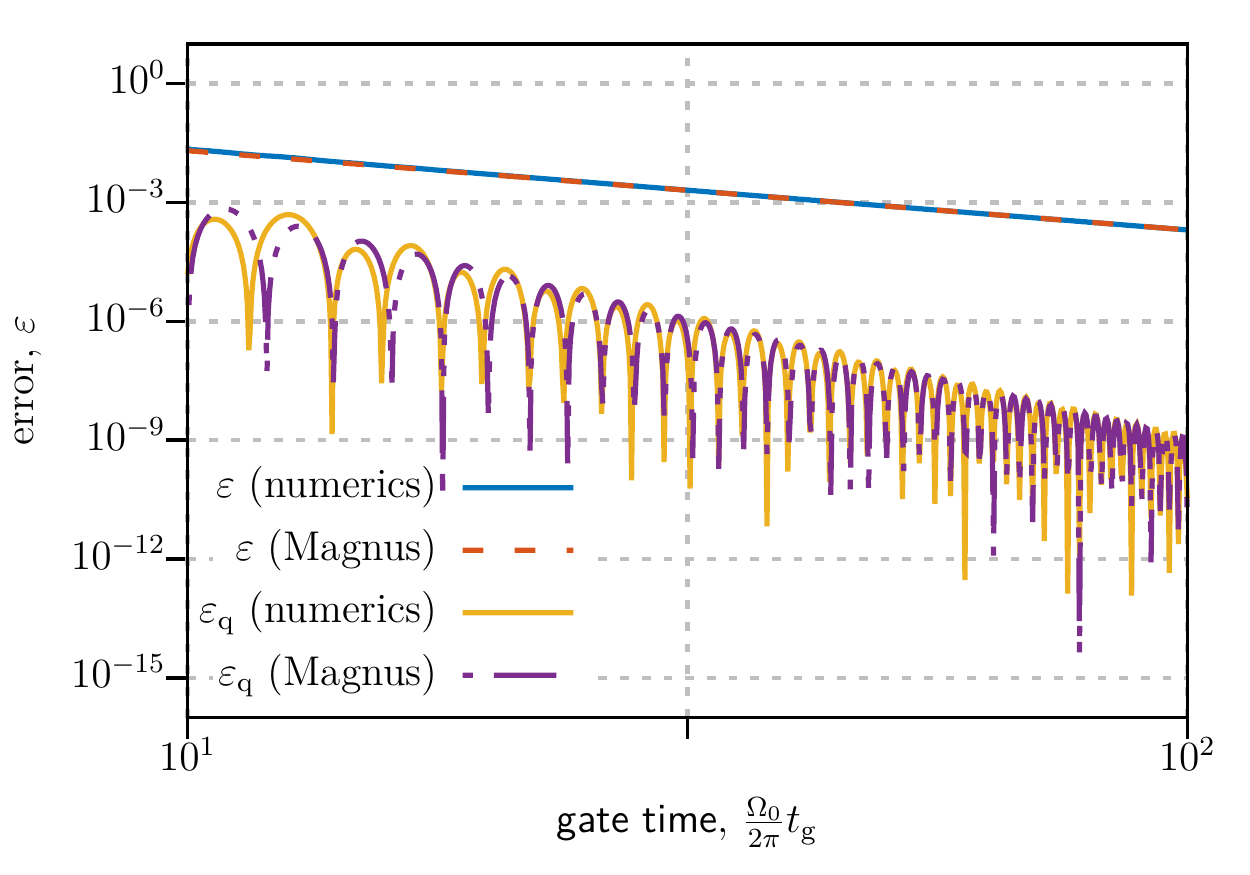}
	\caption{(Color Online) Error $\varepsilon = 1 - \bar{F}$ [see Eq.~\eqref{eq:avgF}] as a function of gate time for the
	unitary operator $\hU_\uG$ acting on both the qubit and auxiliary subspace and its projection $\hU_\uq$ on the qubit
	subspace ($\varepsilon_\uq$). Corrections to the ideal gate due to non-adiabatic transitions only affect the auxiliary subspace at
	leading order in $\hV_\mm{err}$ while the dynamics in the qubit subspace is only affected  at third order. This is a consequence of
	our choice for $\theta (t)$ whose first and second derivates vanish at $t=0$, $t_\ug /2$, and $t_\ug$.}  
	\label{fig:fig02} 
\end{figure}

However, it is misleading to compare $\hU_{\uG,\mm{ad}}$ to $\hU_\uG$ to determine the gating time for which the error
$\varepsilon = 1-\bar{F}$ goes above a critical threshold, e.g. $\varepsilon = 10^{-3}$ for quantum error correction. In
Eq.~\eqref{eq:AvgFidFull} the deviations from unity are only due to imperfect dynamics in the auxiliary subspace as we identified
earlier on [see Eqs.~\eqref{eq:GeomGate} and \eqref{eq:GeomGateNonAdCorr}]. Since Eq.~\eqref{eq:avgF} holds for any linear
operator $\hO$ in a $d$-dimensional Hilbert space, we consider instead $\hO_\uq = \hP_\uq \hU_{\uG,\mm{ad}}^\dag \hP_\uq \hU_\uG
\hP_\uq$ that allows us to quantify errors that only affect the dynamics in the qubit subspace. Here, $\hP_\uq$ is the projector
onto the qubit subspace. $\hO_\uq$ corresponds to measuring the overlap between the unitary operator $\hU_{\uq,\mm{ad}} = \hP_\uq
\hU_{\uG,\mm{ad}} \hP_\uq$ and the operator $\hU_\uq = \hP_\uq \hU_\uG \hP_\uq$; the latter is not necessarily unitary.  Due to
the direct sum structure of $\hU_{\uG,\mm{ad}}$ [cf.~\eqref{eq:GeomGate}], the projection operation yields the ideal gate acting
on the qubit subspace only. 

Within this framework, and performing a fourth order Magnus expansion (see Appendix~\ref{sec:appendixA}), we find 
\begin{equation}
	\bar{F}_\uq = 1 + \frac{a \pi^2}{\left(\Omega_0 t_\ug\right)^6}\left[-1 + \cos\left(\frac{\Omega_0 t_\ug}{4}\right)
	\cos(\gamma_0)\right]\sin^2\left(\frac{\Omega_0 t_\ug}{8}\right)
	\label{eq:avgFqubit}
\end{equation}
with $a = 14 745 600$. We stress that there exists special gate times for which the fidelity is equal to unity. Using
Eq.~\eqref{eq:avgFqubit}, we find those times to be
\begin{equation}
	t_\ug = 
	\begin{dcases}
		&\frac{8 \pi}{\Omega_0} k_1, \quad k_1 \in \mathbb{N}^\ast, \\
		&\frac{4(\pi + 2\pi k_2)}{\Omega_0}, \quad k_2 \in \mathbb{N}.
	\end{dcases}
	\label{eq:AdSpecGateT}
\end{equation}
This phenomenon (a coherent cancellation of non-adiabatic effects) has also been discussed in Ref.~\cite{wiebe2012}.

In Fig.~\ref{fig:fig02}, we plot the error $\varepsilon = 1 - \bar{F}$ [see Eq.~\eqref{eq:avgF}] as a function of gate time for
$\hO$ and $\hO_\uq$ both obtained numerically by integrating the Schrödinger equation and perturbatively via the Magnus expansion.
In the adiabatic regime, i.e. $\Omega_0 t_\ug/2 \to \infty$, the Magnus expansion fully captures deviations from the ideal adiabatic
evolution which confirms our error scaling analysis. However, even with the improved error scaling in the qubit subspace, the
achievable gate times in a realistic setup remain longer than typical decoherence rates.

\section{Accelerated Gate}

We now turn to the main goal of this paper: how can we accelerate the geometric qubit gate presented in
Sec.~\ref{subsec:AdiabaticGate} using the general philosophy of ``shortcuts to adiabaticity''
(STA)~\cite{demirplak2003,demirplak2005,demirplak2008,berry2009,baksic2016}? At first glance, this is a non-trivial problem. The
purpose of STAs is to accelerate the evolution of a {\it specific} initial state, and usually, one does not care about the final,
overall phase of the state. In contrast, we want to accelerate the evolution of an {\it arbitrary} initial qubit state, and the
phases acquired by the adiabatic states are of crucial importance. Despite these difficulties, we show that our desired goal can
indeed be accomplished. We will use our recently proposed dressed-state approach, which allows acceleration of STIRAP-type
processes by simply modifying the form of the original control pulses~\cite{baksic2016,zhou2016}. Note that Ref.~\cite{liu2017}
did not consider these difficulties; as we show below, it is not a trivial task to find a suitable STA that preserves the phases
acquired by the adiabatic states.

Our goal will be to modify the three control fields $\Omega_{i \ue}(t)$ ($i= 0,1,\ua$) from the values given in
Eqs.~\eqref{eq:EnvCtrl} and \eqref{eq:theta}, so that the desired gate operation is accomplished even though the total protocol
time $t_\ug$ is not long compared to $2/\Omega_0$.  This modification can be described by adding a term to the Hamiltonian, i.e.
$\hH (t) \rightarrow \hH (t) + \hW (t) \equiv \hH_\mm{mod} (t)$.  In the adiabatic frame, this modification can be
written as
\begin{equation}
	\begin{aligned}
		\hW_\mm{ad} (t) &= \frac{1}{2} \left[ -W_z (t) \left(\ketbra{\ub_-}{\ub_-} - \ketbra{\ub_+}{\ub_+}\right) \right.\\
		&\phantom{={}}
		+ \frac{W_x (t)}{\sqrt{2}} \left(\ketbra{\ud_2}{\ub_-} + \ketbra{\ud_2}{\ub_+} + \mm{H.c.}\right) \\
		&\phantom{={}}
		\left. + \frac{W_y (t)}{\sqrt{2}}\left(-i \ketbra{\ub_-}{\ud_2} -i \ketbra{\ub_+}{\ud_2} + \mm{H.c.} \right) \right].
	\end{aligned}
	\label{eq:WadGenParam}
\end{equation}
One can readily verify that transforming Eq.~\eqref{eq:WadGenParam} to the original frame, $\hW (t) = \hS_\mm{ad} (t) \hW_\mm{ad}
(t) \hS^\dag_\mm{ad} (t)$, results in a control Hamiltonian having the same form as Eq.~\eqref{eq:HTripod}; no additional control
fields are required. Note that the qubit-only dark state $\ket{\dz}$ remains decoupled for any choice of the $W_j (t)$,
$j\in\{x,y,z\}$.

Conventional STAs attempt to modify $\hH (t)$ so that the evolution follows the (original) adiabatic trajectory at {\it all}
times. In contrast, the dressed-state approach of Ref.~\cite{baksic2016} aims for something less extreme. We let the system
deviate from the adiabatic trajectory at intermediate times. This can be framed as a time-dependent dressing of the original
adiabatic eigenstate, with the dressing vanishing at $t=0$ and $t=t_\ug$. For our problem, we need to add an
additional constraint. We must find a dressed version of the dark state $\ket{\ud_2}$
\begin{equation}
	\ket{\tilde{\ud}_2 (t)} \equiv \hS^\dag_\mu(t) \ket{\ud_2},
    \label{eq:DressCond}
\end{equation}
which retains the geometric nature of the evolution. In other words we require that the new dressed dark state does not acquire any
dynamical phases. 

\subsection{Generic dressing}
\label{sec:generic_dressing}

Following Ref.~\cite{baksic2016} we try a simple dressing transformation
\begin{equation}
	\hS_\mu (t) = \exp\left[ -i \mu (t) \left(\ketbra{\ub_-}{\ud_2} + \mm{H.c.}\right) \right],
	\label{eq:GenDressingLambda1}
\end{equation}
where the dressing angle $\mu (t)$ remains to be determined. It must satisfy $\mu (0) = \mu (t_\ug) = 0$, to ensure the dressing
vanishes at the start and end of the protocol (which then guarantees $\ket{\tilde{\ud}_2(t)} = \ket{\tone}$ at $t=0$ and
$t=t_\ug$). Note that at this stage we only assume $\theta (t)$ to be of the form given in Eq.~\eqref{eq:theta} while no
particular form is assumed for $\gamma (t)$.

The goal is now to pick the dressing parameter $\mu(t)$ and control modifications $\hW (t)$ such that the resulting dynamics
does not cause transitions out of the dressed dark state.  By considering the dynamics in the time-dependent frame defined by
Eq.~\eqref{eq:GenDressingLambda1} (dressing frame), we find that this can be accomplished by choosing the dressing angle to
fulfill the differential equation 
\begin{equation}
	\dot{\mu} (t) = \frac{\sin\left[2 \theta (t)\right]\dot{\gamma}(t)}{\sqrt{2}}
	\label{eq:DressAngle}
\end{equation}
and the control fields to be
\begin{equation}
	\begin{aligned}
		W_x (t) &= \sin[2 \theta (t)]\dot{\gamma} (t),\\
		W_y (t) &= \sqrt{2}\left[ \cos[\theta(t)]^2 \tan[\mu (t)] \dot{\gamma} (t) + \sqrt{2}\dot{\theta} (t) \right], \\
		W_z (t) &= -\Omega_0 + 4\sqrt{2}\cot[2\mu (t)]\dot{\theta} (t) \\
		&\phantom{={}} 
		+\frac{\dot{\gamma} (t)}{2} \left[ 1 + 5 \cos[2\theta(t)]-2\cos[\theta(t)]^2\sec[\mu (t)]^2 \right].
	\end{aligned}
	\label{eq:WctrlsLambda1}
\end{equation}
Unfortunately, suppressing unwanted transitions is not enough to achieve our gate. We also need
control over the {\it phase} acquired by the dressed dark state. In particular, it should not acquire a dynamical phase which
depends explicitly on $t_\ug$ while still accumulating a geometrical phase. We stress that the control fields in
Eq.~\eqref{eq:WctrlsLambda1} also cancel the purely dynamical phase originating from the dressing by ensuring that the energy of
the dressed dark state is $0$. Within this framework the phase accumulated by the dressed dark state is given by
\begin{equation}
	\begin{aligned}
		\varphi_\mm{dds} &= \int_0^{t_\ug} \di{t} \frac{\sec[\mu (t)]^2}{8}\left\{ \dot{\gamma} (t) \left[ 3 + \cos[2\mu (t)]
		\vphantom{\sqrt{2}}\right.\right.\\
		&\phantom{={}}
		\left.\left.- \cos[2\theta(t)] (1+3\cos[2\mu(t)]) \vphantom{\sqrt{2}}\right] + 4\sqrt{2} \sin[2\mu(t)]
		\dot{\theta}(t)\right\}.
	\end{aligned}
	\label{eq:phaseDDS}
\end{equation}
In spite of the similarities with Eq.~\eqref{eq:GeomPhaseGen}, there is no guarantee that this phase is purely geometric since one
might not be able to express $\mu$ as a function of $\gamma$ and $\theta$ only. In our example, however, the situation is far
worse. A solution of Eq.~\eqref{eq:DressAngle} that fulfills the requirement that the dressing must vanish at the boundaries,
[$\mu (0) = \mu (t_\ug) =0$] leads to $\dot{\gamma} (t)$ to be an antisymmetric function on the interval $[0, t_\ug]$ since
$\sin[2 \theta (t)]$ is symmetric on said interval [see Eq.~\eqref{eq:theta}]. Using the symmetry of the functions involved in
Eq.~\eqref{eq:phaseDDS}, one finds $\varphi_\mm{dds} = 0$.

\subsection{Spin-based dressing}

However, for $\gamma (t)$ given by Eq.~\eqref{eq:PhaseAE} and arbitrary $\theta (t)$ of the form in Eq.~\eqref{eq:theta},  one can
find a large class of STAs for which there is an accumulated phase whose nature is geometric.  To proceed, we first define effective
spin-$1$ operators to describe the dressed frame states:  $\hJ_x = (\ketbra{\tilde{\ud}_2}{\tilde{\ub}_+} +
\ketbra{\tilde{\ud}_2}{\tilde{\ub}_-} + \mm{H.c.}) / \sqrt{2}$, $\hJ_y = (i \ketbra{\tilde{\ud}_2}{\tilde{\ub}_-} - i
\ketbra{\tilde{\ud}_2}{\tilde{\ub}_+} + \mm{H.c.})/\sqrt{2}$, and $\hJ_z = (\ketbra{\tilde{\ub}_-}{\tilde{\ub}_-} -
\ketbra{\tilde{\ub}_+}{\tilde{\ub}_+})$.  The dressing transformation of interest is then:
\begin{equation}
	\hS_\nu (t) = \exp\left[-i \nu (t) \hJ_{x,\mm{ad}} \right].
	\label{eq:DressingSx}
\end{equation}
In the dressed frame, the Hamiltonian is given by
\begin{equation}
	\begin{aligned}
		\hH_\mm{dressed} &= \hS_\nu^\dag (t) \hH_\mm{ad} (t) \hS_\nu (t) - i \hS_\nu^\dag (t) \rd_t \hS_\nu^\dag (t) \\
		&= \hH_\mm{spin} (t) + \hH_\mm{ns} (t) + \hH_\mm{geom} (t),
	\end{aligned}
	\label{eq:DressedHGen}
\end{equation}
with
\begin{equation}
	\hH_\mm{spin} (t) = B_z (t) \hJ_z + B_x (t) \hJ_x + B_y (t) \hJ_y
	\label{eq:Hspin}
\end{equation}
\begin{equation}
	\begin{aligned}
	\hH_\mm{ns} (t) &= \frac{\dot{\gamma} (t)}{2}\left[ \Xi_1 (t) \left(\ketbra{\tilde{\ub}_-}{\tilde{\ub}_-} +
	\ketbra{\tilde{\ub}_+}{\tilde{\ub}_+}\right)\right. \\
	&\phantom{={}} 
	+ \Xi_2 (t) \left(\ketbra{\tilde{\ub}_+}{\tilde{\ub}_-} + \mm{H.c.} \right)  \\
	&\phantom{={}} 
	+  \Xi_3 (t) \left(i\ketbra{\tilde{\ub}_+}{\tilde{\ub}_-} + \mm{H.c.} \right) \\
	&\phantom{={}}
	+  \Xi_4 (t) \left(\ketbra{\tilde{\ud}_2}{\tilde{\ub}_-} - \ketbra{\tilde{\ud}_2}{\tilde{\ub}_+ } + \mm{H.c.}\right) \\
	&\phantom{={}} 
	\left. +  \Xi_5 (t) \left( i \ketbra{\tilde{\ud}_2}{\tilde{\ub}_-} +i\ketbra{\tilde{\ud}_2}{\tilde{\ub}_+ } + \mm{H.c.}\right)
	\right]
	\end{aligned}
	\label{eq:HNonSpin}
\end{equation}
and
\begin{equation}
	\hH_\mm{geom} (t) = \dot{\gamma} (t) \sin^2[\theta(t)]\cos^2[\nu (t)] \ketbra{\tilde{\ud}_2}{\tilde{\ud}_2}.
	\label{eq:Hgeom}
\end{equation}
We have written Eq.~\eqref{eq:DressedHGen} as the sum of a spin Hamiltonian [see Eq.~\eqref{eq:Hspin}], a non-spin Hamiltonian
[see Eq.~\eqref{eq:HNonSpin}], and a Hamiltonian that generates the geometric phase [see Eq.~\eqref{eq:Hgeom}].  We also defined
the effective magnetic field components
\begin{equation}
	\begin{aligned}
		B_x (t) &= - \dot{\nu} (t),\\
		B_y (t) &= -\frac{\Omega_0}{2} \sin[\nu (t)] + \dot{\theta} (t) \cos[\nu (t)],\\
		B_z (t) &= -\frac{\Omega_0}{2} \cos[\nu (t)] - \dot{\theta} (t) \sin[\nu (t)],
	\end{aligned}
	\label{eq:BeffHspin}
\end{equation}
as well as the parameters of the non-spin Hamiltonian:
\begin{equation}
	\begin{aligned}
		\Xi_1 (t) &= \cos^2[\theta (t)] + \sin^2[\theta (t)]\sin^2[\nu (t)], \\
		\Xi_2 (t) &= -\cos^2 [\theta (t)] + \sin^2 [\theta (t)]\sin^2 [\nu (t)], \\
		\Xi_3 (t) &= \sin[2 \theta (t)]\sin [\nu (t)], \\
		\Xi_4 (t) &= \frac{\sin[2 \theta (t)] \cos[\nu (t)]}{\sqrt{2}},\\
		\Xi_5 (t) &=  - \frac{\sin^2[\theta (t)] \sin[2 \nu(t)] }{\sqrt{2}}.
	\end{aligned}
	\label{eq:CoeffHns}
\end{equation}

The choice of dressing in Eq.~\eqref{eq:DressingSx} was made to ensure that $\bra{\tilde{\ud}_2}\hH_{\rm dressed}(t) - \hH_{\rm
geom}(t)  \ket{\tilde{\ud}_2} = 0$; this partly solves the problem of the STA giving rise to unwanted dynamical phases in the
evolution of the dressed dark state $\ket{\tilde{\ud}_2}$. In contrast to the STA approach (see Sec.~\ref{sec:generic_dressing}),
we do not start by looking for a dressing angle $\nu(t)$ that generates a dressing transformation that cancels unwanted transitions
(between dressed dark state and dressed bright states).  Instead, we start by looking for a $\nu(t)$ that gives a specific value for the phase accumulated by
the dressed dark state $\ket{\tilde{\ud}_2}$. Neglecting for a moment transitions involving the dressed dark state, we have that
the phase accumulated by $\ket{\tilde{\ud}_2}$ is given by
\begin{equation}
	\varphi_\mm{dds} = \int_0^{t_\ug} \di{t} \sin^2[\theta(t)]\cos^2[\nu (t)] \dot{\gamma} (t).
	\label{eq:GeomPhaseSATD}
\end{equation}
Taking into account that we have chosen a particular form for $\gamma (t)$ [see Eq.~\eqref{eq:PhaseAE}] and comparing
Eq.~\eqref{eq:GeomPhaseSATD} to Eq.~\eqref{eq:GeomPhaseGen}, we see that the phase accumulated by the dressed dark state is equal
to the adiabatic-limit dark state geometric phase $\gamma_0$ if 
\begin{equation}
	\nu(t_\ug/2) = 0,
    \label{eq:DressingFunctionConstraint}
\end{equation}
i.e. we must restrict ourselves to a class of dressing transformations that exactly vanish halfway through the protocol.

There is a second consequence of having to work with dressing transformations that fulfill
Eq.~\eqref{eq:DressingFunctionConstraint}:~one can easily verify that $\hH_\mm{ns} (t)$ does not generate any dynamics. Since
$\dot{\gamma} (t) = \gamma_0 \delta(t-t_\ug/2)$, integrating $\hH_\mm{ns} (t)$ between $t=0$ and $t=t_\ug$ yields $\bm{0}$ because
we evaluate $\hH_\mm{ns} (t)$ at $t=t_\ug/2$ where $\theta (t_\ug/2) = \pi/2$ and $\nu (t_\ug/2)=0$ such that all parameters
defined in Eq.~\eqref{eq:CoeffHns} evaluate to $0$. 

Within this framework all that remains is to find a specific dressing function $\nu(t)$ satisfying
Eq.~\eqref{eq:DressingFunctionConstraint}, and a corresponding control Hamiltonian $\hW (t)$ that cancels unwanted transitions
generated by $\hH_\mm{spin} (t)$.  This is essentially equivalent to the general problem treated in Ref.~\cite{baksic2016}. While
many choices are possible, a particular simple approach is the so-called superadiabatic transitionless driving (SATD) dressing
introduced in Ref.~\cite{baksic2016}.  This is defined by the specific dressing angle 
\begin{equation}
	\nu (t) = \nu_{\rm SATD} (t) = \arctan\left[\frac{2 \dot{\theta} (t)}{\Omega_0}\right]. 
	\label{eq:SATDAngle}
\end{equation}
This satisfies our constraint Eq.~\eqref{eq:DressingFunctionConstraint} as long as the initial pulse sequence satisfies
$\dot{\theta}(t_\ug/2)=0$. For example, the pulse shape in Eq.~\eqref{eq:polyangle} satisfies this property.

Before proceeding we note that the phase accumulated by the dressed dark state can still be viewed as a geometric phase. As long
as Eq.~\eqref{eq:DressingFunctionConstraint} is fulfilled the accumulated phase is independent of the protocol time $t_\ug$ and
does not depend on the details of the pulse. We stress, however, one more time that the dressing transformation allowing one to
get a STA that preserves the geometric nature of the phase accumulated by the dressed dark state $\ket{\tilde{\ud}_2 (t)}$
explicitly depends on our choice of $\gamma (t)$ and that our specific choice of dressing [see Eq.~\eqref{eq:SATDAngle}] further
requires the adiabatic protocol to obey $\dot{\theta}(t_\ug/2)=0$.

For this choice of dressing, the required modified control fields (which cancel transitions out of the dressed dark state) are:
\begin{equation}
	\begin{aligned}
		\Omega_{0\ue} (t) &\to \frac{\Omega_0}{2} \cos(\alpha) \left[\sin[\theta(t)] + 4
		\frac{\cos[\theta(t)]\ddot{\theta}(t)}{\Omega_0^2 +4 \dot{\theta}^2 (t)} \right],\\
		\Omega_{1\ue} (t) &\to \frac{\Omega_0}{2} \sin(\alpha)e^{i\beta} \left[ \sin[\theta(t)] + 4
		\frac{\cos[\theta(t)]\ddot{\theta}(t)}{\Omega_0^2 +4 \dot{\theta}^2 (t)}\right], \\
		\Omega_{\ua \ue} (t) &\to \frac{\Omega_0}{2}e^{i \gamma (t)} \left[\cos[\theta(t)] - 4 \frac{\sin[\theta(t)]
		\ddot{\theta}(t)}{\Omega_0^2 +4 \dot{\theta}^2 (t)} \right].
	\end{aligned}
	\label{eq:ModCtrlFields}
\end{equation}

Combining these results, we find that the accelerated dynamics results in an evolution that at $t=t_\ug$ yields the gate
\begin{equation}
	\hU_{\uG,\mm{SATD}} = e^{-i \frac{\gamma_0}{2}} e^{-i \frac{\gamma_0}{2} \nv \cdot \hat{\sigmav}} \oplus e^{i
	\frac{\gamma_0}{2}} e^{i \frac{\varphi_\mm{SATD}}{2} \nv_{\mm{SATD}} \cdot \hat{\sigmav}_{\ua \ue}},
	\label{eq:STAGate}
\end{equation}
whose action on the qubit subspace is the same as $\hU_{\uG,\mm{ad}}$ [see Eq.~\eqref{eq:GeomGate}], but acts differently on the
auxiliary subspace. The latter undergoes a rotation of angle $\varphi_\mm{SATD} = 2 \arccos[\cos(\gamma_0 /2) \cos(\Phi)]$ around
the axis 
\begin{equation}
	\nv_{\mm{SATD}} =
	\frac{\sin(\Phi)\sin\left(\frac{\gamma_0}{2}\right)}{\sin\left(\frac{\varphi_\mm{SATD}}{2}\right)}\left[-\cot\left(\frac{\gamma_0}{2}\right), 1,
	\cot(\Phi)\right]
	\label{eq:RotAxisAuxSATD}
\end{equation}
with $\Phi = \int_0^{t_\ug /2} \di{t} \sqrt{\Omega_0^2 + 4 \dot{\theta}^2 (t)}$.

We note that Eq.~\eqref{eq:STAGate} always leads to a perfect qubit-subspace fidelity $\bar{F}_q =1$ independent of the speed of the protocol. 

\section{Dissipative Dynamics}

In the following section, we characterize the performance of both the adiabatic and STA gates in the presence of imperfections.
We consider two types of imperfections:~dissipation and uncertainties in the parameters of the Hamiltonian.

To model the loss we consider a Lindblad master equation that describes pure dephasing of the ground and excited states
\begin{equation}	
        \begin{aligned}
                \rd_t \hrho (t) &= -i \left[\hH (t), \hrho (t)\right] \\
                &\phantom{={}}
                + \sum_{j=0,1,\ua,\ue}\Gamma_{\varphi,j} \left[ \ketbra{j}{j} \hrho(t)
                \ketbra{j}{j} - \frac{1}{2} \left\{\ketbra{j}{j}, \hrho (t)\right\}_+ \right],
        \end{aligned}
        \label{eq:Lindblad}
\end{equation}
where $\hH (t)$ is the Hamiltonian, $\hrho (t)$ is the density operator of the system, $\Gamma_{\varphi,j}$ ($j\in
\{0,1,\ua,\ue\}$) is the dephasing rate of state $\ket{j}$, and we have defined the anti-commutator $\{\hO_1, \hO_2\}_+ = \hO_1
\hO_2 + \hO_2 \hO_1$. We stress that in the adiabatic frame the dephasing processes we are considering lead to transitions between
instantaneous eigenstates. For this reason we do not explicitly consider relaxation processes. 

To quantify how decoherence affects the performance of the gate, we use the result of Bowdrey \emph{et al.} for the average
fidelity of single qubit maps~\cite{bowdrey2002}
\begin{equation}
	\bar{F}_\mm{map} = \frac{1}{6} \sum_{j=\pm x, \pm y, \pm z} \mm{Tr} \left[ \hU_\uq \hrho_j \hU_\uq^\dag \hrho_j (t)
	\right],
	\label{eq:AvgFMap}
\end{equation}
where $\hrho_j$ with $j \in  \{\pm x, \pm y, \pm z\}$ is an axial pure state on the Bloch sphere of the qubit, e.g. $\hrho_x = 1/2
(\ket{0} + \ket{1})(\bra{0} + \bra{1})$, $\hU_\uq$ was defined earlier in the text, and $\hrho_j (t)$ is a solution of
Eq.~\eqref{eq:Lindblad} for the initial state $\hrho_j$. 

In addition to errors due to noise, we also consider errors arising from uncertainties in the system Hamiltonian.  Here, we assume
that the amplitude parameter $\Omega_0$ [cf. Eqs.~\eqref{eq:HTripod} and Eq.~\eqref{eq:EnvCtrl}] is only known with finite
precision, as described by a probability distribution $p(\Omega_0)$. In the following we assume $p(\Omega_0)$ to be uniform on the
interval $[\Omega_0 (1-k), \Omega_0 (1+k)]$ with $k \in (0,1)$. The performance of the gate is then quantified via the averaged
average fidelity
\begin{equation}
	\langle \bar{F}_\mm{map} \rangle = \int_{\Omega_0 (1-k)}^{\Omega_0 (1 +k)} \di{\Omega_0}
	p(\Omega_0) \bar{F}_\mm{map} (\Omega_0).
	\label{eq:AvgW0FMap}
\end{equation}

For the results that follow, accelerated gates are implemented using a starting pulse shape given by Eq.~(\ref{eq:polyangle}) and
the SATD STA defined by Eqs.~(\ref{eq:SATDAngle}) and (\ref{eq:ModCtrlFields}).

\subsection{Excited State Dephasing}
\label{sec:ExcStateDeph}

\begin{figure}[t!]
	\includegraphics[width=\columnwidth]{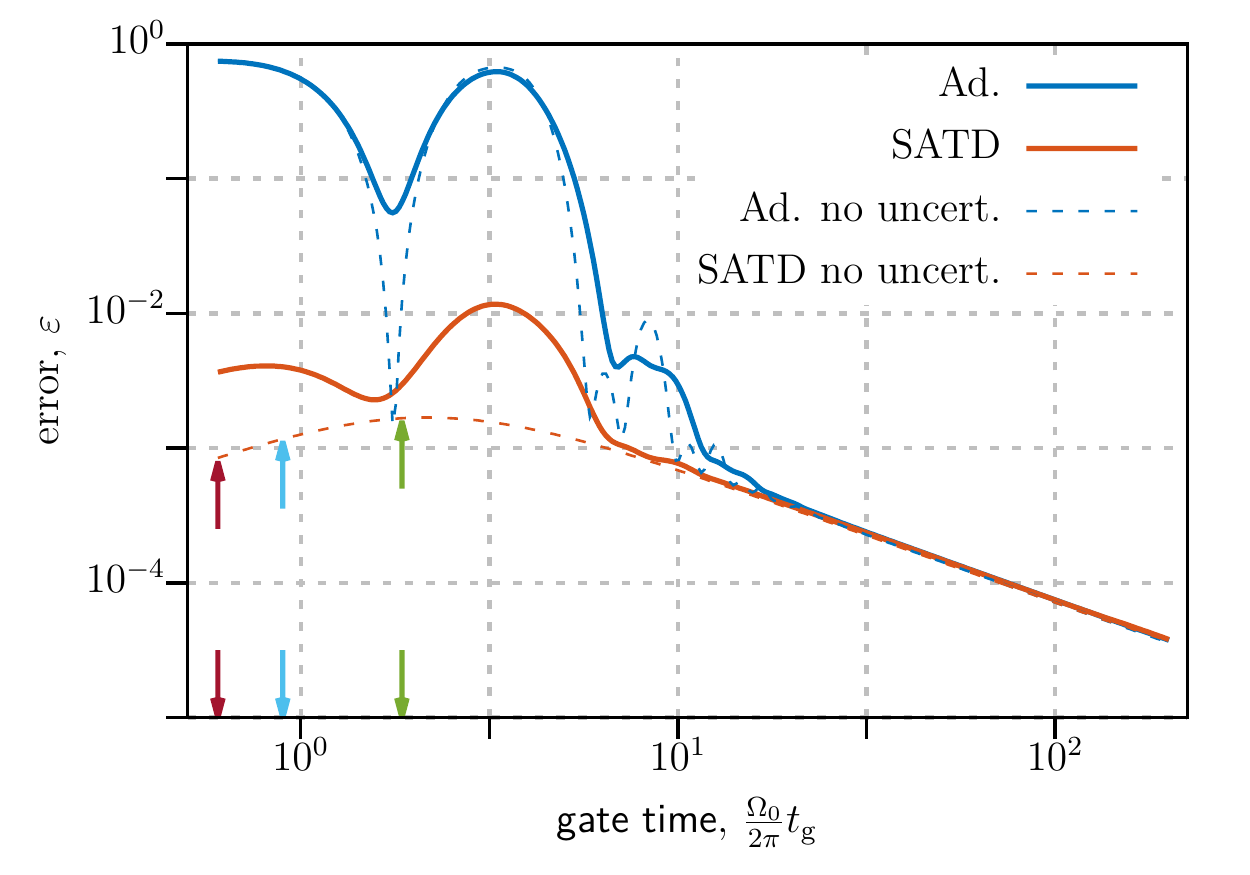}
	\caption{(Color Online) Comparison of the average error $\varepsilon_\mm{map} = 1 - \bar{F}_\mm{map}$ between adiabatic
		(blue traces) and accelerated gate (orange traces) either without uncertainty on $\Omega_0$ (dashed traces) or
		with a $40\%$ ($k=0.2$) uncertainty on $\Omega_0$ (solid traces) as a function of the gate time for $\alpha =
		\pi/4$, $\beta = 0$, $\gamma_0 = \pi$, $\Gamma_{\varphi,\ue} = 10^{-2} (\Omega_0 / 2 \pi)$, and
		$\Gamma_{\varphi,j} = 0$ for $j \in \{0,1,\ua\}$. Shorter gate times than the one indicated by the green arrow 
		result in a modified control sequence whose maximal amplitude is larger than $\Omega_0$. The blue (red) arrow 
		indicates a modified control sequence that requires twice (three times) the energy cost 
		[see Eq.~\eqref{eq:CostProtocol}] of the adiabatic sequence.}
	\label{fig:fig03} 
\end{figure}

In Fig.~\ref{fig:fig03}, we plot the error $\varepsilon_\mm{map} = 1 - \bar{F}_\mm{map}$ as a function of the gate time for
$\alpha = \pi/4$, $\beta = 0$, $\gamma_0 = \pi$, $\Gamma_{\varphi,\ue} = \Omega_0 / (2 \pi \times 100)$, and $\Gamma_{\varphi,j} =
0$ for $j \in \{0,1,\ua\}$ for both the adiabatic (blue traces) and accelerated protocol (orange traces) either without
uncertainty on $\Omega_0$ (dashed traces) or with an uncertainty of $40\%$ ($k=0.2$) on $\Omega_0$ (solid traces). We have
indicated with a green arrow the shortest gate time for which the maximal amplitude of the modified controls is still $\Omega_0$,
with a blue (red) arrow the gate time for which the energy cost to generate the STA is twice (three times) as large as the energy
cost to generate the adiabatic control sequence. We define the energy cost of a control sequence as 
\begin{equation}
	C = \frac{1}{t_\ug} \int_0^{t_\ug} \di{t} \norm{\hH (t)}_2,
	\label{eq:CostProtocol}
\end{equation}
where $\norm{\hH (t)}_2 = \sigma_\mm{max} [\hH (t)]$ is the $p=2$ operator norm equal to the largest singular value of the
Hamiltonian operator $\hH (t)$ denoted by $\sigma_\mm{max} [\hH (t)]$. 

We start by observing that going to the adiabatic regime, $t_\ug \to \infty$, results in both gates becoming insensitive to noise
and uncertainty. This is expected because the evolution of the system can be reduced to the evolution of the dark state
$\ket{\ud_2 (t)}$ [see Eq.~\eqref{eq:dark2}] which contains no excited state amplitude.  We note that for the accelerated gate,
going to the adiabatic regime results in a vanishing dressing such that the dressed dark state is effectively $\ket{\ud_2}$.
Moreover, a small uncertainty on the instantaneous gap is irrelevant as long as $t_\ug \gg 2/\Omega_0$. It is only outside of the
adiabatic regime that both gates are affected by a lossy excited state since during the evolution the excited state will be
occupied. One clearly sees that the accelerated version of the gate outperforms its adiabatic counterpart; this reflects the fact
that the mechanism leading to excited state occupancy is different in each case.

For the adiabatic version of the gate, non-adiabatic processes are responsible for the transitions between the dark state
$\ket{\ud_2}$ and the bright states $\ket{\ub_\pm}$, which contain a finite excited state amplitude [see
Eq.~\eqref{eq:brightSts}]. Since non-adiabatic transitions need first to happen for noise in the excited state to disrupt the
dynamics, the gate error is mainly dominated by non-adiabatic errors. However, signatures of the dissipative dynamics can be
observed for the special times [see Eq.~\eqref{eq:AdSpecGateT}] for which the coherent evolution brings the system back to the
dark state. These special times do not lead to a perfect gate anymore because the coherent mechanism that brings the system back
to the dark state is disrupted by excited state dephasing. This coherent mechanism is further hindered by the uncertainty on
$\Omega_0$.

The accelerated gate is constructed such that whatever amplitude leaves the dark state it has to come back to the dark state by
the end of the protocol; this is equivalent to having the system remaining in the dressed dark state $\ket{\tilde{\ud}_2 (t)}$ for
the whole evolution. However, leaving the dark state is not harmless in the presence of a noisy excited state:~it disrupts the STA
dynamics in two ways. First, there is no guarantee that the amplitude leaving the dark state comes back by the end of the
evolution. Second, even if the amplitude that left the dark state comes back, it could come back with a phase error. These two
mechanism can be identified as the leading order processes leading to deviations of the average fidelity [Eq.~\eqref{eq:AvgFMap}]
from unity. Using a Magnus expansion we can find approximate solutions of Eq.~\eqref{eq:Lindblad} which we use to evaluate
$\bar{F}_\mm{map}$ [Eq.~\eqref{eq:AvgFMap}] (see Appendix~\ref{sec:appendixB}). We find 
\begin{equation}
	\begin{aligned}
		\bar{F}_\mm{map} &= 1 - \frac{4}{3} \Gamma_{\varphi,\ue} \int_0^{t_\ug /2} \di{t} \frac{\dot{\theta}^2
		(t)}{\Omega_0^2 + 4 \dot{\theta}^2 (t)} \\
		&\phantom{={}} 
		- \frac{8}{3} \Gamma_{\varphi,\ue} \Omega_0^2 \int_0^{t_\ug /2} \di{t}
		\frac{\dot{\theta}^2 (t)}{\left[\Omega_0^2 + 4 \dot{\theta}^2 (t)\right]^2} \\
		&\phantom{={}} 
		+ \mathcal{O}\left[\left(\frac{\Gamma_{\varphi,\ue}}{\Omega_0^2 t_\ug}\right)^2\right],
	\end{aligned}
	\label{eq:AvgFMapExc}
\end{equation}
which is in good agreement with numerical simulations [see Appendix~\ref{sec:appendixB}]. In particular, Eq.~\eqref{eq:AvgFMapExc}
captures the non-monotonic behavior of the gate error as a function of gate time. Sufficiently faster gates become insensitive to
a lossy excited state. We note, however, that reaching such a regime experimentally is difficult. 

In the presence of uncertainty in $\Omega_0$, it is impossible to realize the exact STA that would cancel out non-adiabatic
transitions. As a result, a small non-adiabatic transition probability from the dressed dark state to the bright states remains.
This becomes apparent for faster gate times where the fidelity of the accelerated gates oscillates in sync with the fidelity of
the adiabatic gate.

Finally, we note that in this scenario if being fast is not essential, then there is no benefit in using an accelerated gate. Both
gates perform equally well in the adiabatic regime. However, if the gate time has to be below a threshold outside of the adiabatic
regime, then the accelerated gate will outperformed the adiabatic one. 

\subsection{Ground and Excited State Dephasing}
\label{sec:AllDeph}

\begin{figure}[t!]
	\includegraphics[width=\columnwidth]{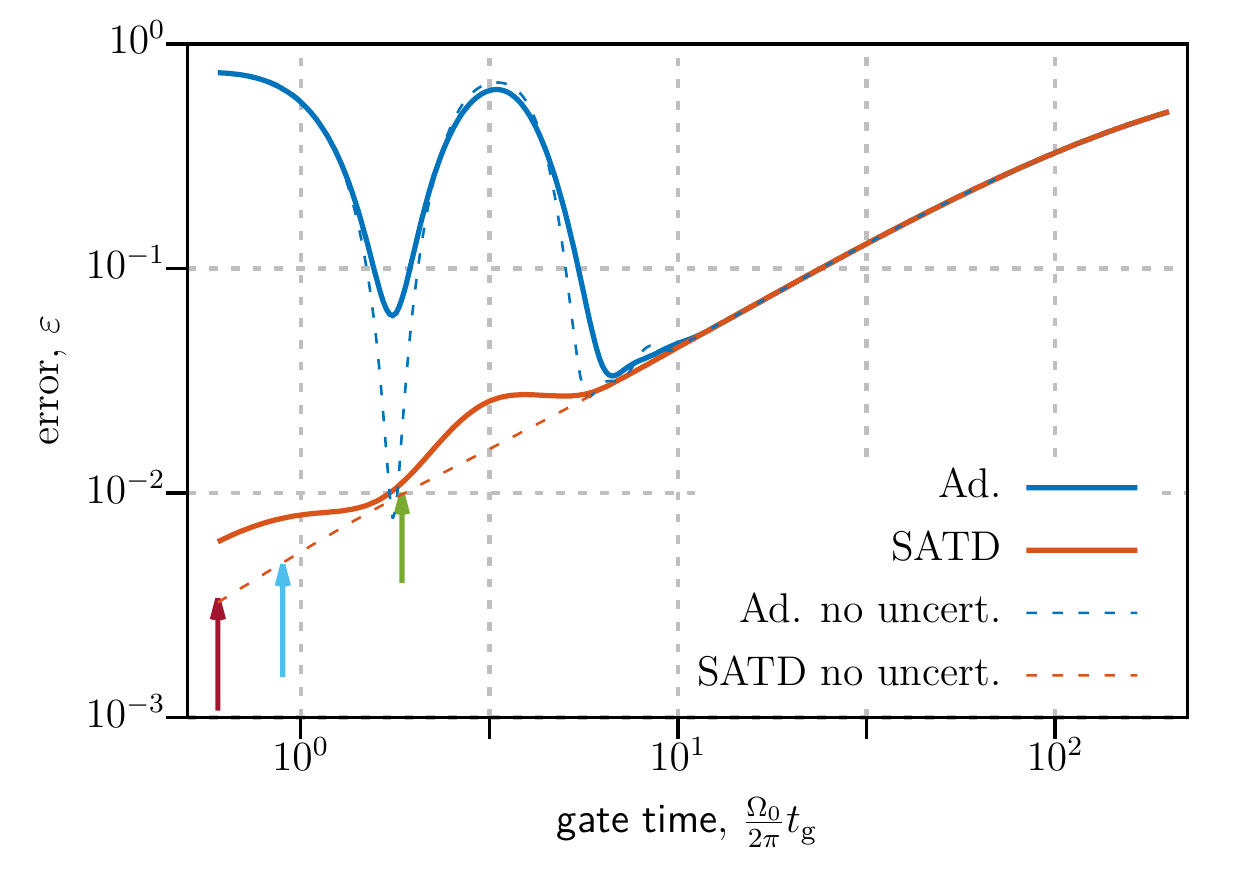}
	\caption{(Color Online) Comparison of the average error $\varepsilon_\mm{map} = 1 - \bar{F}_\mm{map}$ between adiabatic
		(blue traces) and accelerated gate (orange traces) either without uncertainty on $\Omega_0$ (dashed traces) or
		with a $40\%$ ($k=0.2$) uncertainty on $\Omega_0$ (solid traces) as a function of the gate time for $\alpha =
		\pi/4$, $\beta = 0$, $\gamma_0 = \pi$, $\Gamma_{\varphi,\ue} = 10^{-2} (\Omega_0 / 2 \pi )$, and
		$\Gamma_{\varphi,j} = 10^{-2} (\Omega_0 / 2 \pi )$ for $j \in \{0,1,\ua\}$.  Shorter gate times than the one
		indicated by the green arrow result in a modified control sequence whose maximal amplitude is larger than
		$\Omega_0$. The blue (red) arrow indicates a modified control sequence that requires twice (three times) the energy
		cost [see Eq.~\eqref{eq:CostProtocol}] of the adiabatic sequence.}
	\label{fig:fig04} 
\end{figure}

In Fig.~\ref{fig:fig04}, we plot the error, $\varepsilon_\mm{map} = 1 - \bar{F}_\mm{map}$, as a function of the gate time for
$\alpha = \pi/4$, $\beta = 0$, $\gamma_0 = \pi$, $\Gamma_{\varphi,\ue} = \Omega_0 / (2 \pi \times 100)$, and
$\Gamma_{\varphi,0}=\Gamma_{\varphi,1}=\Gamma_{\varphi,\ua}=\Omega_0 / (2 \pi \times 100)$ for both the adiabatic (blue traces)
and accelerated protocol (orange traces) either without uncertainty on $\Omega_0$ (dashed traces) or with an uncertainty of $40\%$
($k=0.2$) on $\Omega_0$ (solid traces). Similarly to Fig.~\ref{fig:fig03}, we have indicated with a green arrow the shortest gate
time for which the maximal amplitude of the modified controls is still $\Omega_0$, with a blue (red) arrow the gate time for which
the energy cost [see Eq.~\eqref{eq:CostProtocol}] to generate the STA is twice (three times) as large as the energy cost to
generate the adiabatic control sequence. 

In contrast to the case where only the excited state is lossy, operating in the adiabatic regime does not lead to a perfect gate.
The dephasing on the ground state manifold sets a threshold for the slowest ``allowed'' gate time. As a consequence, the adiabatic
version of the gate becomes an unviable option. Trying to perform the gate faster to avoid ground state dephasing unavoidably
leads to a regime where the dominating source of errors are non-adiabatic transitions. On the other hand, the accelerated gate is
less susceptible to non-adiabatic errors. As a result decreasing the gate time allows one to escape the interval for which the
error is mainly dominated by ground state dephasing, $t_\ug \ll 1/(\Gamma_{\varphi,0} + \Gamma_{\varphi,1} +
\Gamma_{\varphi,\ua})$, to operate in a regime where only excited state dephasing contributes to the gate error. 

In this scenario, being fast becomes essential and corresponds to a situation where the accelerated gate provides a real benefit
over its adiabatic counterpart. 

\subsection{Extended Robustness Comparison}

\begin{figure}[t!]
	\includegraphics[width=\columnwidth]{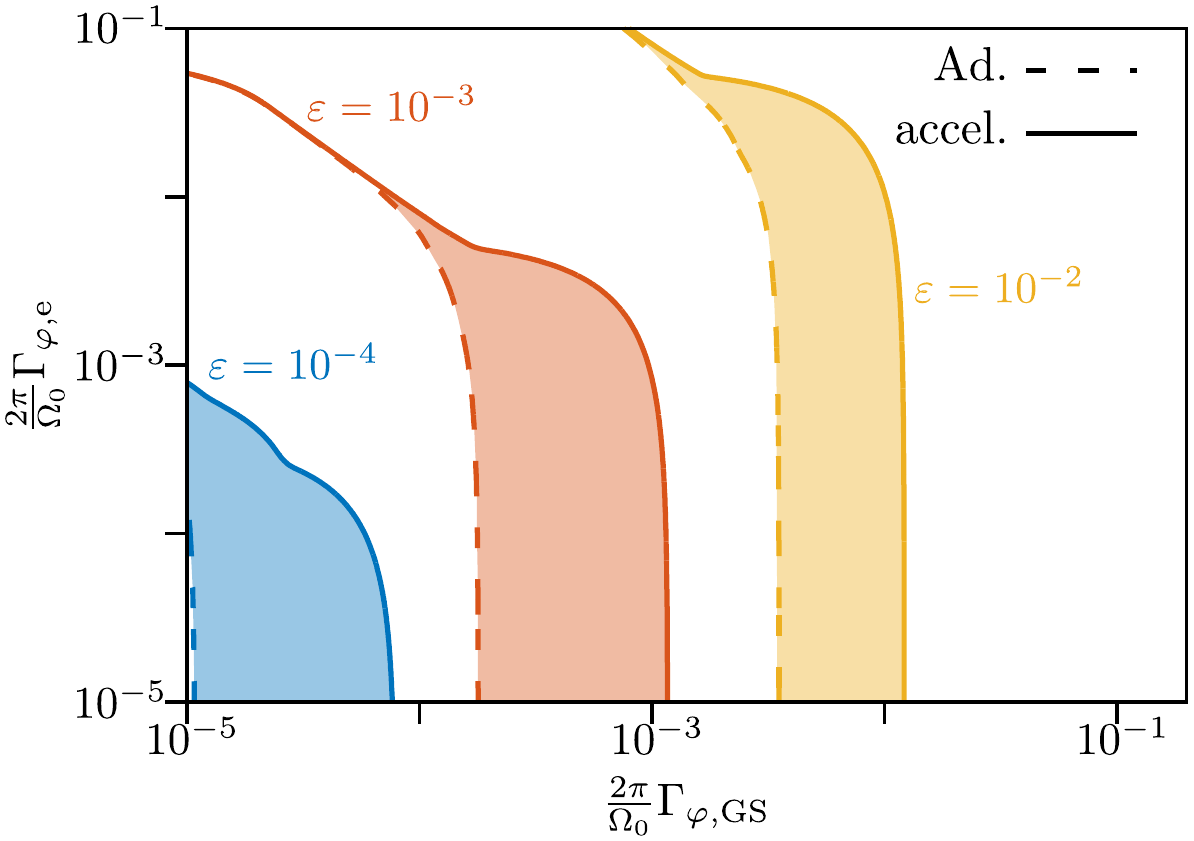}
	\caption{(Color Online) Comparison of the smallest gate error between adiabatic and accelerated gate. For fixed values of
	the dephasing rates we look for the gate time that yields the smallest error. We consider an uncertainty on $\Omega_0$
	of $10\%$ ($k=0.05$) and $\Gamma_{\varphi,0} = \Gamma_{\varphi,1} = \Gamma_{\varphi,\ua} = \Gamma_{\varphi,\mm{GS}}$. The
	accelerated gate reaches the same error as the adiabatic gate for dephasing rates that are roughly one order of magnitude 
	larger.}
	\label{fig:fig05} 
\end{figure}

To identify the regimes where the accelerated gate provides a clear benefit over the adiabatic gate, we look for the gate time
that yields the smallest error for fixed dephasing rates. We have constrained the minimal gate time by imposing that the maximal
amplitude of the modified pulse sequence cannot be larger than $\Omega_0$. For simplicity we consider the case where the ground
state dephasing rates are equal, i.e. $\Gamma_{\varphi,0} = \Gamma_{\varphi,1} = \Gamma_{\varphi,\ua} = \Gamma_{\varphi,\mm{GS}}$.
We also assume that the uncertainty on $\Omega_0$ is $10\%$ ($k=0.05$). In Fig.~\ref{fig:fig05}, we plot contour lines for
different error thresholds as a function of $\Gamma_{\varphi,\mm{GS}}$ and $\Gamma_{\varphi,\ue}$ for the adiabatic gate (dashed
lines) and accelerated gate (solid lines). For the displayed contours, we see that the accelerated gate reaches the same error
level as the adiabatic gate for rates that can be roughly up to one order of magnitude larger. 

\section{Conclusion}

We have shown how to use the framework of shortcuts-to-adiabaticity to accelerate geometric gates in tripod systems.  We have
discussed both how standard STA techniques designed for the transfer of a single known state are problematic due to STA-induced
modification of dynamical and geometric phases. We have also shown a set of protocols that overcome this seeming limitation and
discussed the advantages of using accelerated gates in the presence of dissipation and Hamiltonian uncertainties:~the accelerated
gate preserves the robustness against parameter variations and allows one to be fast enough to overcome thresholds set by
relaxation and dephasing times. Our accelerate control scheme can be implemented in a variety of state-of-the-art qubit
implementations. 

Our work also suggests that accelerated geometric-based two-qubit gates could be developed for a variety of systems. In
particular, it could greatly benefit superconducting-based architectures where two-qubit gates are still the main limitation
preventing the realization of high-fidelity complex gate sequences.

\section{Acknowledgements}

This work was supported by the Army Research Office under Grant No. W911NF-19-1-0328.

\begin{appendix}

\section{Average Gate Fidelity with Unitary Evolution}
\label{sec:appendixA}

In this appendix we show how to use a Magnus expansion to obtain approximate solutions of the Schrödinger equation 
\begin{equation}
	i \rd_t \hU (t) = \hH (t) \hU (t),
	\label{eq:SchroedingerEq}
\end{equation}
where $\hH (t)$ was defined in Eq.~\eqref{eq:HLambda} of the main text. We focus on the special case where the
control field phase $\gamma (t)$ is given by Eq.~\eqref{eq:PhaseAE}. The special form of $\gamma (t)$ and the symmetry of the
function $\theta (t)$ [see Eq.~\eqref{eq:theta}] allows us to split the evolution into two distinct STIRAP processes defined by
the Hamiltonians
\begin{equation}
	\hH_1 (t) = \frac{\Omega_0}{2}\left[ \sin[\thetat (t)] \ketbra{\tone}{\ue} + \cos[\thetat (t)] \ketbra{\ua}{\ue} +
		\mm{H.c.}
	\right],
	\label{eq:STIRAP1}
\end{equation}
which describes the first half of the evolution that brings the system from $\ket{\tone}$ to $\ket{\ua}$, and
\begin{equation}
	\hH_2 (t) = \frac{\Omega_0}{2}\left[ \cos[\thetat (t)] \ketbra{\tone}{\ue} + e^{i \gamma_0} \sin[\thetat (t)]
	\ketbra{\ua}{\ue} + \mm{H.c.} \right],
	\label{eq:STIRAP2}
\end{equation}
which describes the second half of the evolution that brings $\ket{\ua}$ back to $\ket{\tone}$. Here, $\thetat (t) = \pi P(t) /2$
[see Eq.~\eqref{eq:polyangle}] is defined for $t \in [0, t_\ug/2]$ and we have used the symmetry of the function $\theta (t)$ [see
Eq.~\eqref{eq:theta}] to obtain Eq.~\eqref{eq:STIRAP2}. Within this framework the evolution operator $\hU$ can be parametrized as 
\begin{equation}
	\hU (t) = 
	\begin{cases}
		\hU_1 (t) &0 \leq t < \frac{t_\ug}{2}, \\
		\hU_2 (t - t_\ug /2) \hU_1 (t_\ug /2) & \frac{t_\ug}{2} \leq t < t_\ug,
	\end{cases}
	\label{eq:UMagnusGen}
\end{equation}
where $\hU_1 (t)$ is generated by $\hH_1 (t)$ [see Eq.~\eqref{eq:STIRAP1}] and $\hU_2 (t)$ is generated by $\hH_2 (t)$ [see
Eq.~\eqref{eq:STIRAP2}]. We stress that $\hU (t)$ is continuous at $t=t_\ug/2$ because the actual coupling strength between
$\ket{\ua}$ and $\ket{\ue}$ is 0, which allows us in the first place to have a phase jump.

It is useful to look for solutions of Eq.~\eqref{eq:SchroedingerEq} in the adiabatic frame.  We can transform
Eqs.~\eqref{eq:STIRAP1}  and \eqref{eq:STIRAP2} to the adiabatic frame by using the frame-change operator $\hS_\mm{ad} (t)$
defined in Eq.~\eqref{eq:SadLambda} of the main text with 
\begin{equation}
	\ket{\ud_2 (t)} =
	\begin{dcases}
		\cos[\thetat (t)]\ket{\tone} - \sin[\thetat (t)] \ket{\ua} &\text{for }\hH_1 (t) \\
		\sin[\thetat (t)]\ket{\tone} - e^{i \gamma_0} \cos[\thetat (t)] \ket{\ua} &\text{for } \hH_2 (t)
	\end{dcases}
	\label{eq:dark2Appendix}
\end{equation}
and 
\begin{equation}
	\ket{\ub_\pm (t)} =
	\begin{dcases}
		&\frac{1}{\sqrt{2}} \left( \pm \sin[\thetat (t)]\ket{\tone} \pm \cos[\thetat (t)]
		\ket{\ua} + \ket{\ue} \right)\\ 
		&\text{for } \hH_1 (t), \\
		&\frac{1}{\sqrt{2}} \left( \pm \cos[\thetat (t)]\ket{\tone} \pm e^{i \gamma_0} \sin[\thetat (t)]
		\ket{\ua} + \ket{\ue} \right)\\ 
		&\text{for } \hH_2 (t).
	\end{dcases}
	\label{eq:brightAppendix}
\end{equation}
We find 
\begin{equation}
	\hH_\mm{ad,1(2)} (t) = -\frac{\Omega_0}{2}\hJ_{z,\mm{ad}} \pm \thetatd (t) \hJ_{y,\mm{ad}},
	\label{eq:HadAppendix}
\end{equation}
where $\hH_\mm{ad,1}$ ($\hH_\mm{ad,2}$) denotes $\hH_1 (t)$ [$\hH_2 (t)$] in the adiabatic frame. We have introduced the spin
operators $\hJ_{z,\mm{ad}} = (\ketbra{\ub_-}{\ub_-} - \ketbra{\ub_+}{\ub_+})$ and $\hJ_{y,\mm{ad}} = (i \ketbra{\ud_2}{\ub_+} - i
\ketbra{\ud_2}{\ub_-} + \mm{H.c.})/\sqrt{2}$. We note that the remaining spin operator $\hJ_{x,\mm{ad}}$ was introduced in the main
text below Eq.~\eqref{eq:DressingSx}. It is convenient to transform Eq.~\eqref{eq:HadAppendix} to the interaction picture
generated by $\hH_{0,\mm{ad}} = -\Omega_0 \hJ_{z,\mm{ad}} /2$ to perform the Magnus expansion, this yields
\begin{equation}
	\begin{aligned}
	\hV_{\uI,1(2)} (t) &= \mp i e^{-i \frac{\Omega_0}{2} t} \frac{\thetatd (t)}{\sqrt{2}}\ketbra{\ub_-}{\ud_2} \pm i
		e^{i \frac{\Omega_0}{2} t} \frac{\thetatd(t)}{\sqrt{2}}\ketbra{\ub_+}{\ud_2} \\
		&\phantom{={}}+ \mm{H.c.}.
	\end{aligned}
	\label{eq:VIad}
\end{equation}
It is useful to notice that the dynamics generated by $\hV_{\uI,1(2)} (t)$ can be parametrized as 
\begin{equation}
	\begin{aligned}
		&\hU_{\uI,1(2)} (t) = \exp\left[-i \left(\Delta (t) \hJ_{z,\mm{ad}} \pm \Omega_x (t) \hJ_{x,\mm{ad}} \pm \Omega_y (t)
		\hJ_{y,\mm{ad}}\right) \right] \\
		&= \exp\left[-i \xi (t)  \left(n_z (t) \hJ_{z,\mm{ad}} \pm n_x (t) (t) \hJ_{x,\mm{ad}} \pm n_y (t)
		\hJ_{y,\mm{ad}}\right) \right] 
	\end{aligned}
	\label{eq:Ugen}
\end{equation}
with $\xi (t) = \sqrt{\Delta^2 (t) + \Omega_x^2 (t) + \Omega_y^2 (t)}$, $n_x (t) = \Omega_x (t)/\xi (t)$, $n_y (t) = \Omega_y (t)
/\xi (t)$, and $n_z (t) = \Delta(t) /\xi (t)$. This form allows us to get an exact representation for $\hU_{\uI,1(2)} (t)$ by
expanding the exponential into a series and using the properties of the spin operators. The functions $\Delta (t)$, $\Omega_x
(t)$, and $\Omega_y (t)$ are found perturbatively using a fourth-order Magnus expansion~\cite{magnus1954}. We find that at
$t=t_\ug/2$ these functions evaluate to 
\begin{equation}
	\begin{aligned}
		\Delta (t_\ug/2) &= \frac{a_1}{\Omega_0 t_\ug} + \frac{a_2}{\left(\Omega_0 t_\ug \right)^3}, \\
		\Omega_x (t_\ug/2) &= b_1 \frac{\sin^2\left(\frac{\Omega_0 t_\ug}{8}\right)}{\left(\Omega_0 t_\ug\right)^3}
		+b_2 \frac{\sin\left(\frac{\Omega_0 t_\ug}{4}\right)}{\left(\Omega_0 t_\ug\right)^4}, \\
		\Omega_y (t_\ug/2) &= c_1 \frac{\sin\left(\frac{\Omega_0 t_\ug}{4}\right)}{\left(\Omega_0 t_\ug\right)^3} + c_2
		\frac{\cos^2\left(\frac{\Omega_0 t_\ug}{8}\right)}{\left(\Omega_0 t_\ug\right)^4},
	\end{aligned}
	\label{eq:MagnusFields}
\end{equation}
with $a_1=-5\pi^2/7$, $a_2 = 4500 \pi^4/2431 - 960\pi^2/7$, $b_1 = 3840 \pi$, $b_2 = 960 \pi (336+5\pi^2)/7$, $c_1 = -1920\pi$,
and $c_2 = -1920\pi (336 + 5\pi^2)/7$.

Finally, the evolution operators are given by 
\begin{equation}
	\hU_{1(2)} (t) = \hS_{\mm{ad},1(2)} \hU_0 (t) \hU_{\uI,1(2)} (t) \hS_{\mm{ad,1(2)}}^\dag (0),
	\label{eq:UFinal}
\end{equation}
with $\hU_0 (t) = \exp[i \Omega_0 t \hS_{z,\mm{ad}} /2]$. 

We can evaluate Eq.~\eqref{eq:avgF} with $\hO = \hO_\uq$ [see text below Eq.~\eqref{eq:AvgFidFull}] and $\hU_\uG = \hU (t_\ug)$
[see Eq.~\eqref{eq:UMagnusGen}] (not shown due to the length of the result). To get Eq.~\eqref{eq:avgFqubit}, one further needs to
expand the trigonometric functions involved in the result to fourth-order in $\xi (t)$ and collect terms up to sixth-order in
$1/(\Omega_0 t_\ug)$.

\section{Average Map Fidelity with Excited State Dephasing}
\label{sec:appendixB}

\begin{figure}[t!]
	\includegraphics[width=\columnwidth]{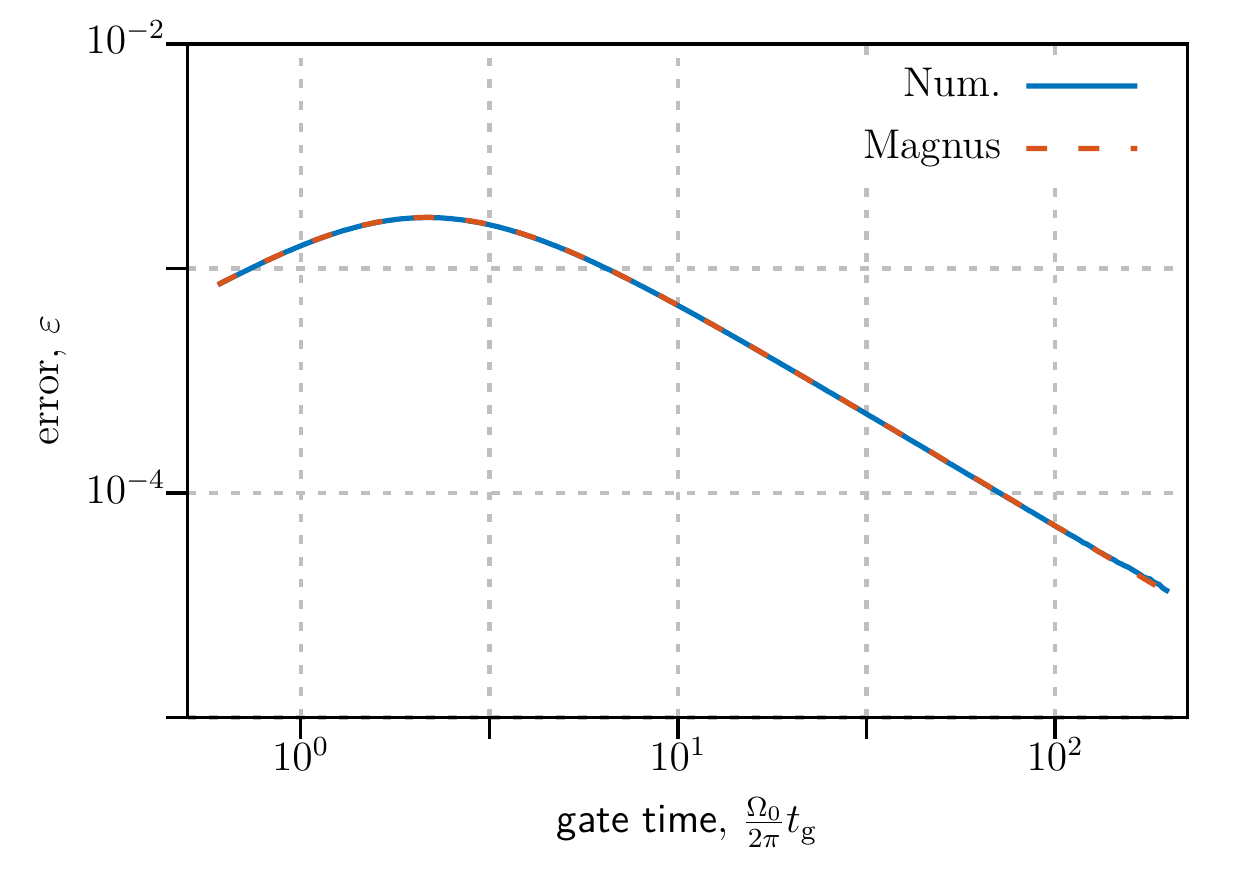}
	\caption{(Color Online) Comparison of the average error $\varepsilon_\mm{map} = 1 - \bar{F}_\mm{map}$ calculated
		numerically (solid blue trace) and using Eq.~\eqref{eq:AvgFMapExc} (dashed orange trace). We used 
		$\alpha = \pi/4$, $\beta = 0$, $\gamma_0 = \pi$, $\Gamma_{\varphi,\ue} = \Omega_0 / (2 \pi \times 100)$.}
	\label{fig:fig_appB} 
\end{figure}

In this section we present the general framework allowing us to evaluate perturbatively the average fidelity of the qubit map [cf.
Eq.~\eqref{eq:AvgFMap}] for the accelerated gate in the presence of excited state dephasing. We start by defining the modified
Hamiltonian with the SATD correction
\begin{equation}
	\hH_\mm{mod} (t) = \hH (t) + \hW_\mm{SATD} (t), 
	\label{eq:HmodDeph}
\end{equation}
where $\hH (t)$ is the Hamiltonian of the tripod system written in terms of the new qubit states [cf. Eq.~\eqref{eq:HLambda}] and 
\begin{equation}
	\begin{aligned}
		\hW_\mm{SATD} &= \frac{2 \Omega_0 \ddot{\theta} (t)}{\Omega_0^2 + 4 \dot{\theta}^2 (t)} \times\\
		&\phantom{={}}\left[\cos[\theta (t)]
		\ketbra{\tone}{\ue} - e^{i \gamma(t)} \sin[\theta(t)] \ketbra{\ua}{\ue} + \mm{H.c.}\right].
		\end{aligned}
	\label{eq:WSATD}
\end{equation}

In the frame defined by the SATD dressing [see Eqs.~\eqref{eq:DressingSx} and \eqref{eq:SATDAngle}], the master equation
describing the evolution of the tripod system with excited dephasing is given by 
\begin{equation}
	\begin{aligned}
	\rd_t \hrho_\mm{dr} (t) &= -i \left[\hH_\mm{mod,dr} (t), \hrho_\mm{dr} (t) \right] \\
	&\phantom{={}}
	+ \Gamma_{\varphi,\ue} \sum_{\substack{i,j,k,l=\\ \tilde{\ub}_-, \tilde{\ub}_+, \tilde{\ud}_2}} 
	c_i (t) c_j^\ast (t) \ketbra{i}{j} \hrho_\mm{dr} (t) c_k (t) c_l^\ast (t) \ketbra{k}{l} \\
	&\phantom{={}}
	-\frac{1}{2} \Gamma_{\varphi,\ue} \sum_{\substack{i,j= \\ \tilde{\ub}_, \tilde{\ub}_+, \tilde{\ud}_2}} \left\{c_i (t)
	c_j^\ast (t) \ketbra{i}{j},\hrho_\mm{dr} (t)\right\}_+
	\end{aligned}
	\label{eq:LindbladDressed}
\end{equation}
with $c_{\tilde{\ub}_-} (t) = c_{\tilde{\ub}_+} (t) = 1/\left(\sqrt{2} \sqrt{1 + 4\dot{\theta}^2 (t)/\Omega_0^2}\right)$ and
$c_{\tilde{\ud_2}} = 2 i \dot{\theta} (t) /\left(\Omega_0 \sqrt{1+ 4\dot{\theta}^2 (t)/\Omega_0^2}\right)$.

Using a superoperator formalism Eq.~\eqref{eq:LindbladDressed} can be written as 
\begin{equation}
	\rd_t \hrho_\mm{dr} (t) = \left[\pmb{\ell}_0 (t) + \pmb{\ell}_\varphi (t)\right] \hrho_\mm{dr},
	\label{eq:SOLindbladDressed}
\end{equation}
where $\pmb{\ell}_0 (t) = i \left[\hH_\mm{mod,dr}^\intercal (t) \otimes \mathbbm{1} - \mathbbm{1} \otimes \hH_\mm{mod,dr}
(t)\right]$ and  $\pmb{\ell}_\varphi=$ $\hL^\ast (t) \otimes \hL (t)$ $- (1/2) \mathbbm{1} \otimes \hL^\intercal (t) \hL (t)$ $
-(1/2) \hL^\intercal (t) \hL^\ast (t) \otimes \mathbbm{1}$ and we have defined $\hL_\varphi (t) = \sqrt{\Gamma_{\varphi,\ue}}
\sum_{i,j} c_i (t) c_j^\ast (t) \ketbra{i}{j}$ with $i,\,j \in \{\tilde{\ub}_-, \tilde{\ub}_+, \tilde{\ud}_2\}$. We also
introduced the complex conjugation and transpose operation denoted by $\hL^\ast (t)$ and $\hL^\intercal (t)$, respectively.

To find approximate solutions of Eq.~\eqref{eq:SOLindbladDressed} it is convenient to work in the interaction picture defined by
$\pmb{\ell}_0 (t)$. In this frame Eq.~\eqref{eq:SOLindbladDressed} reduces to 
\begin{equation}
	\rd_t \hrho_\mm{dr,\uI} (t) = \pmb{\ell}_{\varphi,\uI} (t) \hrho_\mm{dr,\uI} (t),
	\label{eq:SOLindbladDressedInt}
\end{equation}
where $\pmb{\ell}_{\varphi,\uI} (t) = \mathcal{L}_0^\dag (t) \pmb{\ell}_\varphi (t) \mathcal{L}_0 (t)$ and $\mathcal{L}_0 (t)$ is
the solution of $\rd_t \hrho_\mm{dr} (t) = \pmb{\ell}_0 (t) \hrho_\mm{dr} (t)$, i.e. $\hrho_\mm{dr} (t) = \mathcal{L}_0 (t)
\hrho_\mm{dr} (0)$. Within this framework the solution of Eq.~\eqref{eq:SOLindbladDressed} is given by $\hrho_\mm{dr} (t) =
\mathcal{L}_0 (t) \mathcal{L}_\uI (t) \hrho_\mm{dr} (0)$ where  $\mathcal{L}_\uI (t)$ is a solution of
Eq.~\eqref{eq:SOLindbladDressedInt}. Using a first order Magnus expansion~\cite{magnus1954}, we can approximate  $\mathcal{L}_\uI
(t)$ by
\begin{equation}
	\begin{aligned}
		\mathcal{L}_\uI (t) & \simeq \exp\left[ \int_0^t \di{t}_1 \pmb{\ell}_{\varphi,\uI} (t_1)  \right] \\
		&\simeq \mathbbm{1} + \int_0^t \di{t}_1 \pmb{\ell}_{\varphi,\uI} (t_1).
	\end{aligned}
	\label{eq:MagnusSOInt}
\end{equation}
This leads to an approximate solution for $\hrho_\mm{dr} (t)$,
\begin{equation}
	\hrho_\mm{dr} (t) \simeq \mathcal{L}_0 (t) \left[\mathbbm{1} + \int_0^t \di{t}_1 \pmb{\ell}_{\varphi,\uI} (t_1)\right],
	\label{eq:ApproxRhoDiss}
\end{equation}
which can be used to evaluate Eq.~\eqref{eq:AvgFMap}. 

In Fig.~\ref{fig:fig_appB} we plot the error $\varepsilon_\mm{map}$ as a function of gate time for $\alpha=\pi/4$, $\beta=0$,
$\gamma_0 = \pi$, and $\Gamma_{\varphi,\ue} = \Omega_0/(2\pi \times 100)$ for the accelerated protocol calculated numerically
(solid blue trace) and using Eq.~\eqref{eq:AvgFMapExc} (dashed orange trace). As stated in the main text, the approximate
analytical result is in very good agreement with the numerical results.

\end{appendix}

\end{document}